\documentclass[review]{interact}
\usepackage{xcolor}
\usepackage[version=4]{mhchem}
\usepackage{epstopdf}% To incorporate .eps illustrations using PDFLaTeX, etc.
\usepackage[caption=false]{subfig}% Support for small, `sub' figures and tables

\usepackage[numbers,sort&compress,merge]{natbib}% Citation support using natbib.sty
\bibpunct[, ]{[}{]}{,}{n}{,}{,}% Citation support using natbib.sty
\renewcommand\bibfont{\fontsize{10}{12}\selectfont}% Bibliography support using natbib.sty

\theoremstyle{plain}% Theorem-like structures provided by amsthm.sty

\theoremstyle{definition}

\theoremstyle{remark}

\begin{document}

%\articletype{New Views}% Specify the article type or omit as appropriate

\title{Three-body recombination in physical chemistry}

\author{
\name{M. Mirahmadi\textsuperscript{a} and J. P\'erez-R\'ios\textsuperscript{b,c}\thanks{CONTACT J. P\'erez-R\'ios. Email: jesus.perezrios@stonybrook.edu} }
\affil{\textsuperscript{a}Fritz-Haber-Institut der Max-Planck-Gesellschaft, Faradayweg 4-6, D-14195 Berlin}
\affil{\textsuperscript{b} Department of Physics and Astronomy, Stony Brook University, Stony Brook, NY 11794, USA}
\affil{\textsuperscript{c}Institute for Advanced Computational Science, Stony Brook University, Stony Brook, NY 11794, USA}
}

\maketitle

\begin{abstract}

Three-body recombination, or ternary association, is a termolecular reaction in which three particles collide, forming a bound state between two, whereas the third escapes freely. Three-body recombination reactions play a significant role in many systems relevant to physics and chemistry. In particular, they are relevant in cold and ultracold chemistry, quantum gases, astrochemistry, atmospheric physics, physical chemistry, and plasma physics. As a result, three-body recombination has been the subject of extensive work during the last 50 years, although primarily from an experimental perspective. Indeed, a general theory for three-body recombination remains elusive despite the available experimental information. Our group recently developed a direct approach based on classical trajectory calculations in hyperspherical coordinates for three-body recombination to amend this situation, leading to a first principle explanation of ion-atom-atom and atom-atom-atom three-body recombination processes. This review aims to summarize our findings on three-body recombination reactions and identify the remaining challenges in the field.

\end{abstract}


\centerline{\bfseries Contents}\vspace{-14pt}\hfill {\sc{page}}\medskip

\hspace*{-12pt} {\bf{{1.}    Introduction}}\hfill {\pageref{intro}}\\
{\bf{{2.}    Direct three-body recombination}}\hfill {\pageref{direct}}\\
\hspace*{10pt}{2.1.}  Classical trajectory method in hyperspherical coordinates\hfill {\pageref{method}}\\
\hspace*{10pt}{2.2.}  Hyperspherical coordinates\hfill {\pageref{hspher}}\\
\hspace*{10pt}{2.3.}  Scattering in 6D\hfill {\pageref{subsec:sigma}}\\
\hspace*{10pt}{2.4.}  Computational details\hfill {\pageref{subsec:comput}}\\
\hspace*{10pt}{2.5.} Grand angular momentum\hfill {\pageref{subsec:GAM}}\\
{\bf{{3.}    Threshold laws for three-body recombination reactions}}\hfill {\pageref{threshold}}\\
{\bf{{4.}    Universality in few-body processes: hyperradial distribution}}\hfill {\pageref{b3b}}\\
{\bf{{5.}    Atom-atom-atom three-body recombination reactions}}\hfill {\pageref{aaa}}\\
{\bf{{6.}    Ion-atom-atom three-body recombination reactions}}\hfill {\pageref{cnn}}\\
\hspace*{10pt}{6.1.}   Cold Chemistry\hfill {\pageref{coldchemistry}}\\
\hspace*{10pt}{6.2.}   Hyper-thermal chemistry\hfill {\pageref{hyper}}\\
\hspace*{10pt}{6.3.}   Plasma physics\hfill {\pageref{plasma}}\\
{\bf{{7.}  Three-body recombination including internal degrees of freedom}}\hfill {\pageref{degreesoffreedom}} \\
\hspace*{10pt}{7.1.}   Molecule-atom-atom three-body recombination \hfill {\pageref{maa}}\\
\hspace*{24pt} {7.1.1.}   Ozone formation\hfill {\pageref{ozone}}\\
\hspace*{24pt} {7.1.2.}  Charged-neutral-neutral systems\hfill {\pageref{mionatomatom}}\\
\hspace*{10pt}{7.2.}   Molecule-molecule-atom three-body recombination \hfill {\pageref{mma}}\\
\hspace*{10pt}{7.3.}   Molecule-molecule-molecule three-body recombination \hfill {\pageref{mmm}}\\
{\bf{{8.}   Outlook and future challenges}}\hfill {\pageref{outlook}}\\

\section{Introduction}\label{intro}
\begin{figure}[h]
	\centering
	\includegraphics[width=0.6\linewidth]{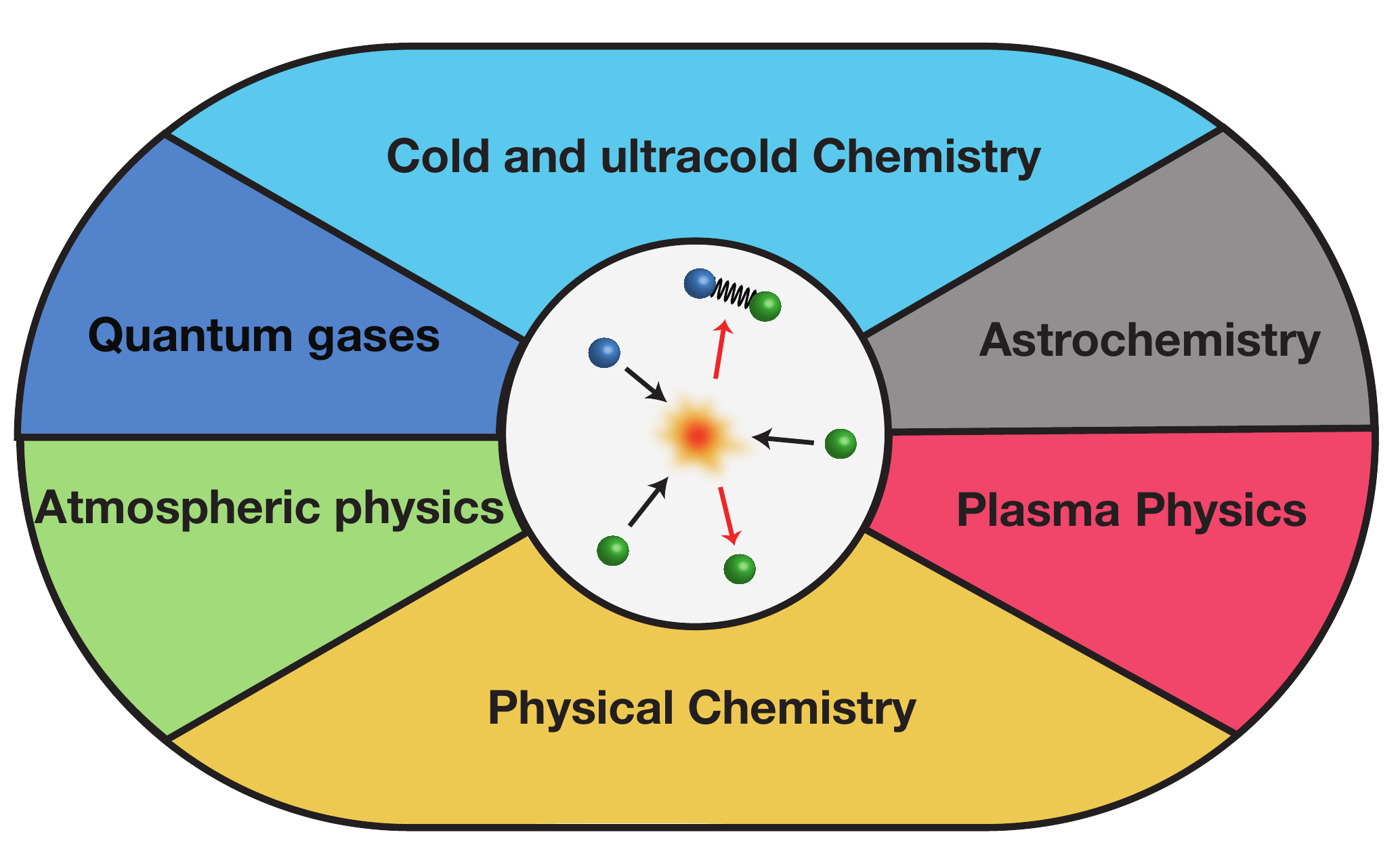}
 	\caption{Three-body recombination reactions in physics and chemistry.}
 	\label{fig1}
\end{figure}

Three-body recombination or ternary association is a termolecular reaction such that three particles collide, forming a bound state between two of them, whereas the third remains free. Despite being less frequent than two-body collisions, three-body recombination processes are relevant to many areas of physics and chemistry, as it is sketeched in Fig.~\ref{fig1}. For instance, in cold and ultracold systems, three-body recombination affects the stability of quantum gases~\cite{Esry1999,Weiner1999,Bedaque2000,Suno2003,Weber2003,Schmidt2020,Koehler2006,Blume2012}, controls the evolution of an ion in a cold dense media~\cite{Krukow2016,Perez-Rios2021,Perez-Rios2020,Mohammadi2021,Weckesser2021}, is essential for anions in a buffer gas~\cite{Hauser2015,Mikosch2008,Wild2021}, or explains the formation of van der Waals molecules in buffer gas cells~\cite{Brahms2008,Suno2009,Brahms2010,Brahms2011,Wang2011,Tariq2013,Quiros2017,Mirahmadi2021,Mirahmadi2021a}. In atmospheric physics, three-body recombination is responsible for the primary reaction toward ozone formation, i.e., O + O$_2$ +M $\rightarrow$ O$_3$ + M, where M=Ar, N$_2$ or O$_2$~\cite{Charlo2004,Luther2005,Kaufmann2006,Mirahmadi2022,SCH06:625,TEP16:19194}. In astrophysics, hydrogen recombination reaction, H + H + H $\rightarrow$ H$_2$ + H, is essential to explain star formation in protoplanetary regions~\cite{Palla1983,Flower2007,Turk2011,Forrey2013}. Similarly, in plasma physics, three-body recombination may affect the divertor detachment physics~\cite{Krsti2003,Cretu2022,Fletcher2007}.

\begin{figure}
	\centering
	\includegraphics[width=0.6\linewidth]{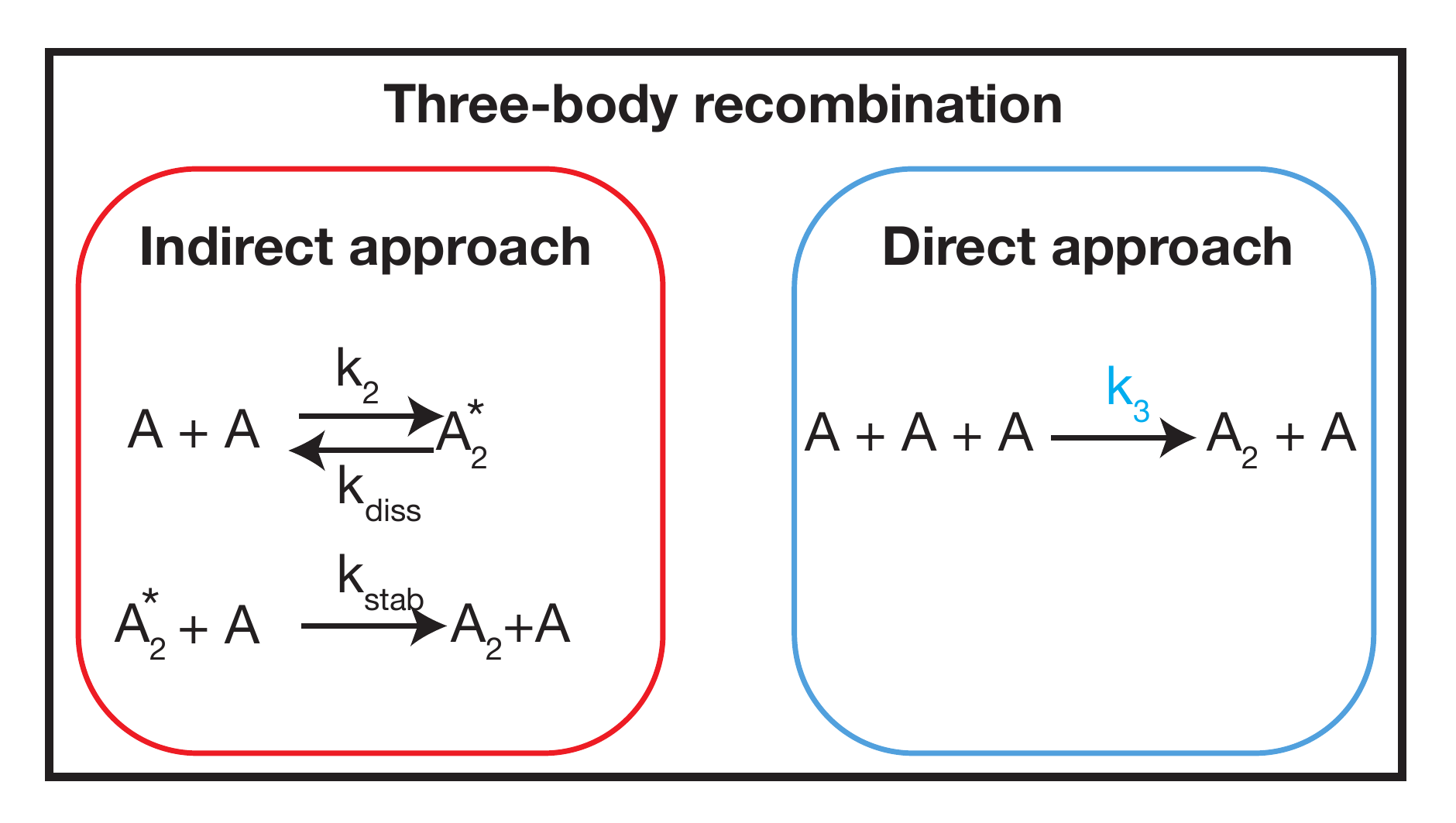}
 	\caption{The two approaches to three-body recombination reactions: the indirect approach based on a two-step reaction mechanism or the direct approach without invoking the presence of an intermediate complex.}
 	\label{fig2}
\end{figure}

On the theory front, three-body recombination has been generally treated following the Lindemann-Hinshelwood mechanism, labeled here as the indirect approach and sketched in Fig.~\ref{fig2}. Within this theoretical framework, two bodies collide to form an intermediate complex before a third body comes into play to stabilize the complex, leading to a new bound state. For instance, in the case of atomic three-body recombination, i.e., \ce{ A + A  + A \rightarrow A$_2$ + A }, known as the Roberts- Bernstein-Curtiss mechanism~\cite{RBCmechanism,RBCmechanism2}, three-body recombination could occur via

\begin{align}\label{eq:2step}
\ce{ A + A &<-->[k_2][k_{diss}] A$_2$^* }\\
\ce{ A$_2$^* + A &->[k_{stab}] A$_2$ + A },
\end{align}
where $k_2$ denotes the rate of formation of A$_2^*$ complexes, $k_{\text{diss}}$ stands for its dissociation rate and $k_{\text{stab}}$ refers to the stabilization rate due to a collision with a third body. Assuming that the production of complexes reaches a steady state, one finds that the three-body recombination rate is given by

\begin{equation}\label{eq7}
    k_3=\frac{k_2k_{\text{stab}}}{k_\text{diss}+k_{\text{stab}}[\text{A}]},
\end{equation}
where [A] is the number density of particle A. Therefore, the nature of the intermediate complex and the atomic number density establishes the outcome of the reaction. Based on this assumption, different theoretical approaches have been explicitly developed using the microscopic reversibility principle, leading to statistical and state-specific phase space theories~\cite{Herbst1979,Herbst1979bisbis,Herbst1980,Bass1979,Bass1981,Bates1979bis,Bates1985,Bates1988,Jennings1982,Ferguson1985}. These theories have a similar theoretical framework but differ in the number and nature of states available for the intermediate complex. As a result, conveying a specific understanding of a given reaction has been possible. However, a general theory that explains every three-body recombination reaction is still lacking. 

On the other hand, an alternative approach to treating termolecular reactions is the direct approach, which is formally equivalent to the indirect one~\cite{Forrey2015}. That is, the computation of the reaction rate without invoking intermediate complex formation, as shown in Fig.~\ref{fig2},

\begin{align}
\ce{ A + A +A ->[k_3] A$_2$ + A }.
\end{align}
\noindent
In this case, the three-body recombination rate only depends on the underlying potential energy surface, the mass of the atoms and the temperature. This approach has been successfully applied into atomic three-body recombination in the ultracold regime via solving the three-body Schr\"odinger equation via the adiabatic hyperspherical method~\cite{Greene2017,Naidon2017}, Fadeev equations or using effective field theory~\cite{Braaten2006,Hammer2010,Naidon2017}. However, these approaches require model potentials with the right two-body scattering length but supporting only a handful of bound states. Therefore, it is challenging to develop a fully quantum mechanical treatment of three-body recombination reactions. Although, it is possible by using classical trajectory calculations. Indeed, using the pioneering ideas of Smith and Shui~\cite{Smith1962,Smith1960,Johnson1980,Johnson1983,Shui,Shui2}, we have been able to develop a general classical trajectory method in hyperspherical coordinates to treat direct atomic three-body recombination processes involving ions and atoms~\cite{Perez-Rios2014,Perez-Rios2015,Perez-Rios2018,Perez-Rios2020,Perez-Rios2021,Mirahmadi2021,Mirahmadi2021a,Mirahmadi2022,Mohammadi2021,Krukow2016,Yu2023,Mirahmadi2023}. In particular, we have studied van der Waals molecule formation in buffer gas cells, ion-atom-atom three-body recombination from the cold to the hyper-thermal regime and ozone formation reaction. Additionally, we have derived threshold laws that have been experimentally confirmed in the case of ion-atom-atom. Similarly, we have identified short-range interactions' role in atomic three-body recombination and the relevance of accurate potential energy surfaces.  

This review presents a method mainly developed in our group to treat direct three-body recombination reactions and the main results obtained in recent years. In particular, we summarize the chief results on van der Waals molecule formation, cold chemistry, hyper-thermal chemistry, and plasma physics regarding atomic and ion-atom-atom three-body recombination. In the same vein, we discuss our results on ozone formation reaction. Finally, we present an overview of experimentally reported termolecular reactions but still poorly understood and how our approach could shed some light on them.

\section{Direct three-body recombination reactions}\label{direct}

This section introduces a methodology to treat direct three-body recombination reactions based on classical trajectory calculations in hyperspherical coordinates. This approach has been successfully applied to different reactions: ion-atom-atom three-body recombination, ozone formation reaction, and atomic three-body recombination relevant to van der Waals molecule formation.

\subsection{Classical trajectory method in hyperspherical coordinates}\label{method}
\begin{figure}[h]
	\begin{center}
		\includegraphics[scale=0.2]{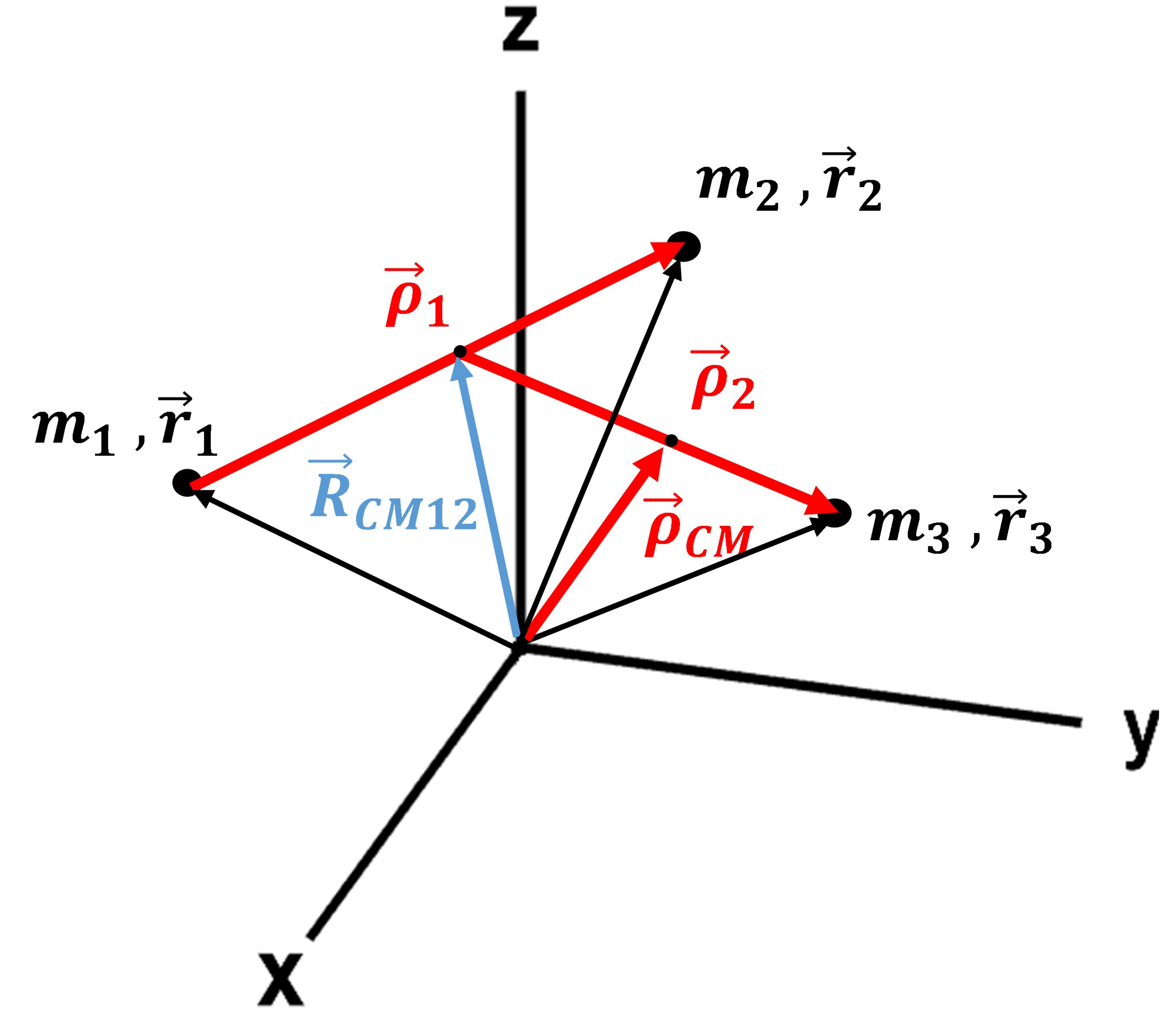}
		\caption{Jacobi coordinates for the three-body problem. The red vectors illustrate the Jacobi vectors, the black arrows indicate the position of the three particles in Cartesian coordinates, and the blue arrow indicates the two-body center-of-mass vector $\vec{R}_{\text{CM}12}$.}	
		\label{fig:jacobi}	
	\end{center}
\end{figure}

The general Hamiltonian governing the classical dynamics of three particles with masses $m_i$ and position vectors $\vec{r}_i$ ($i = 1,2,3$), interacting via the potential $V(\vec{r}_1,\vec{r}_2,\vec{r}_3)$, is given by
%~\footnote{Please note that the same applied to a quantum system in which the momentum is an operator.} 
\begin{equation}\label{eq:cartesianH}
	H = \frac{\vec{p}_1^{~2}}{2m_1} + \frac{\vec{p}_2^{~2}}{2m_2} + \frac{\vec{p}_3^{~2}}{2m_3} + V(\vec{r}_1,\vec{r}_2,\vec{r}_3) ~,
\end{equation}
with  $\vec{p}_i$ being the momentum vector of the $i$-th particle. It is more convenient to consider such systems in Jacobi coordinates~\cite{Pollard1976,suzuki1998}, defined by the following relations	
\begin{align}\label{eq:jacobitrans}
	\vec{\rho}_1 &= \vec{r}_2 - \vec{r}_1 ~, \nonumber \\
	\vec{\rho}_2 &= \vec{r}_3 -\vec{R}_{CM12} ~, \nonumber \\
	\vec{\rho}_{CM} &= \frac{m_1\vec{r}_1 + m_2\vec{r}_2 + m_3\vec{r}_3}{M} ~,
\end{align}
where  $M = m_1 + m_2 + m_3$ is the total mass and $ \vec{R}_{CM12} = (m_1\vec{r}_1 + m_2\vec{r}_2)/(m_1+m_2)$  and $\vec{\rho}_{CM}$ are the center-of-mass vectors of the two-body and three-body systems, respectively. 
The Jacobi vectors are illustrated in Fig.~\ref{fig:jacobi}.

Since the total linear momentum is conserved, the center of mass degrees of freedom can be neglected and the Hamiltonian~\eqref{eq:cartesianH} will be written as
\begin{equation}\label{eq:jacobiH}
	H = \frac{\vec{P}_1^2}{2\mu_{12}} + \frac{\vec{P}_2^2}{2\mu_{3,12}} +  V(\vec{\rho}_1,\vec{\rho}_2) ~,
\end{equation} 
where $\mu_{12}=m_1m_2/(m_1 + m_2)$ and $\mu_{3,12}=m_3(m_1+m_2)/M$ are the associated reduced masses and $\vec{P}_1$ and $\vec{P}_2$ indicate the conjugated momenta of Jacobi vectors $\vec{\rho}_1$ and $\vec{\rho}_2$, respectively. Note that $V( \vec{\rho}_1,\vec{\rho}_2)$ is the same potential as in Eq.~(\ref{eq:cartesianH}), expressed in terms of the Jacobi vectors.

Knowing that the Hamilton's equations of motion are invariant under the (canonical) transformation given by Eq.~(\ref{eq:jacobitrans}), we can find the classical trajectories in terms of Jacobi coordinates by solving ($i=1,2$):
\begin{align}\label{eq:Heqns}
	\frac{d\vec{\rho}_i}{d t} = \frac{\partial H}{\partial \vec{P}_i} ~, \quad \quad 
	\frac{d\vec{P}_i}{d t} = -\frac{\partial H}{\partial \vec{\rho}_i} ~,
\end{align}

\subsection{Hyperspherical coordinates}\label{hspher}
Classically, the N-body scattering problem in three-dimensional (3D) space can be considered as the scattering problem of a single particle with a definite momentum moving towards a scattering center in a (3N-3)-dimensional space by mapping the independent relative coordinates of the N-body system into the (3N-3) degrees of freedom of the single particle. Similarly, one can define the initial conditions as single entities in a higher dimensional space and avoid the complications in the calculations caused by relative coordinates in the 3D space. Therefore, studying the three-body recombination process as a scattering problem in a six-dimensional (6D) space is convenient. Here, the 6D space is characterized by hyperspherical coordinates consisting of five hyperangles $\alpha_j$ ($j = 1,2,3,4,5$) and the hyperradius $R$. Components of the vector $(x_1,x_2,x_3,x_4,x_5,x_6)$, describing a point in the 6D space, can be written as~\cite{Lin1995,Avery2012}  
\begin{align}\label{eq:hsCart}
	x_1 & = R \sin(\alpha_1)\sin(\alpha_2)\sin(\alpha_3)\sin(\alpha_4)\sin(\alpha_5) ~, \nonumber \\
	x_2 & = R  \cos(\alpha_1)\sin(\alpha_2)\sin(\alpha_3)\sin(\alpha_4)\sin(\alpha_5) ~, \nonumber \\
	x_3 & = R  \cos(\alpha_2)\sin(\alpha_3)\sin(\alpha_4)\sin(\alpha_5) ~, \nonumber \\
	x_4 & = R  \cos(\alpha_3)\sin(\alpha_4)\sin(\alpha_5) ~, \nonumber \\
	x_5 & = R  \cos(\alpha_4)\sin(\alpha_5) ~, \nonumber \\
	x_6 & = R  \cos(\alpha_5) ~,
\end{align} 
where $0\leq\alpha_1<2\pi$ and $0\leq\alpha_{j>1}\leq\pi$ and the volume element in the hyperspherical coordinate system is given by 
\begin{align}\label{eq:hsvol}
	d\tau &=  R^5d R\prod_{j=1}^{5}\sin^{j-1}(\alpha_j)d\alpha_j \nonumber \\
	&= R^5d Rd\Omega  ~.
\end{align} 

\subsection{Scattering in 6D}\label{subsec:sigma}
Position and momentum vectors in a 6D space can be constructed from Jacobi vectors and their conjugated momenta as~\cite{Perez-Rios2014,Perez-Rios2020}
\begin{equation}\label{eq:rho6D}
	\vec{\rho} = \begin{pmatrix} \vec{\rho}_1 \\ \vec{\rho}_2 \end{pmatrix} ~,
\end{equation}
and 
\begin{equation}\label{eq:P6D}
	\vec{P} = \begin{pmatrix} \sqrt{\frac{\mu}{\mu_{12}}}\vec{P}_1 \\ \sqrt{\frac{\mu}{\mu_{3,12}}}\vec{P}_2 \end{pmatrix} ~,
\end{equation}
respectively, with $\mu = \sqrt{m_1 m_2 m_3/ M}$ being the three-body reduced mass. Using these definitions and Eq.~(\ref{eq:jacobiH}), we can write the Hamiltonian in the 6D space as
\begin{equation}\label{eq:6DH}
	H^\mathrm{6D} = \dfrac{\vec{P}^2}{2\mu} + V(\vec{\rho}) ~.
\end{equation}

The cross section of the (classical) scattering in the regular 3D space is defined as the area drawn in a plane perpendicular to the particle's initial momentum, which the particle's trajectory should cross to be scattered (i.e., deviation from the uniform rectilinear motion). This concept can be extended to the 6D space as an area in a five-dimensional hyperplane, embedded in the 6D space, perpendicular to the initial momentum vector $\vec{P}_0$. In a similar vein, we define the impact parameter vector $\vec{b}$ as the projection of the initial position vector $\vec{\rho}_0$ on this hyperplane ($\vec{b}\cdot\vec{P}_0 = 0$).

As it has been mentioned earlier, considering the three-body collision as a scattering problem of a single particle in a 6D space, it is possible to uniquely define the initial conditions and the impact parameter as single entities. Therefore, the probability of a three-body recombination event is given as a function of the impact parameter $\vec{b}$ and the initial momentum $\vec{P}_0$ in the hyperspherical coordinate system. Consequently, the total cross section of the three-body recombination process for a particular initial collision energy $E_c=P_0^2/(2\mu)$ is given by (averaging over different orientations of $\vec{P}_0$)
\begin{align}\label{eq:sigma}
	\sigma_\mathrm{rec}(E_c) & = \dfrac{\int \mathcal{P}(\vec{P}_0,\vec{b})  b^4 d b ~d\Omega_bd\Omega_{P_0}}{\int d\Omega_{P_0}} \nonumber \\
	&= \frac{8\pi^2}{3}\int_{0}^{b_\mathrm{max}(E_c)} \mathcal{P}(E_c,b)  b^4 d b ~,
\end{align}
Here, $d\Omega_b$ and $d\Omega_{P_0}$ are the differential elements of the solid hyperangles associated with vectors $\vec{b}$ and $\vec{P}_0$, respectively [see Eq.~(\ref{eq:hsvol})]. 
The function $\mathcal{P}$ is the so-called opacity function and indicates the probability of a recombination event as a function of the impact parameter and collision energy. The angular dependence of the opacity function $\mathcal{P}(\vec{P}_0,\vec{b})$, which depends on both the direction and magnitude of impact parameter and initial momentum vectors, has been averaged out by making use of the Monte Carlo (MC) sampling of trajectories. $b_\mathrm{max}$ in Eq.~(\ref{eq:sigma}) represents the largest impact parameter for which three-body recombination occurs, or in other words, for $b>b_\mathrm{max}$ we have $\mathcal{P}(E_c,b) = 0$. 

Finally, the energy-dependent three-body recombination rate will be written as
\begin{equation}\label{eq:k3}
	k_3(E_c) = \sqrt{\frac{2E_c}{\mu}}\sigma_\mathrm{rec}(E_c) ~.
\end{equation}
Using the appropriate three-body Maxwell-Boltzmann distribution of collision energies, we obtain the corresponding thermal average via ($k_B$ is the Boltzmann constant)
\begin{align}\label{eq:MBLANg}
	k_3(T) &=  \frac{1}{2(k_BT)^3} \int_{0}^{\infty}k_3(E_c) E_c^2 e^{-E_c/(k_BT)} dE_c ~.
\end{align}

\subsection{Computational details}\label{subsec:comput}
\begin{figure}
	\begin{center}
		\includegraphics[scale=0.35]{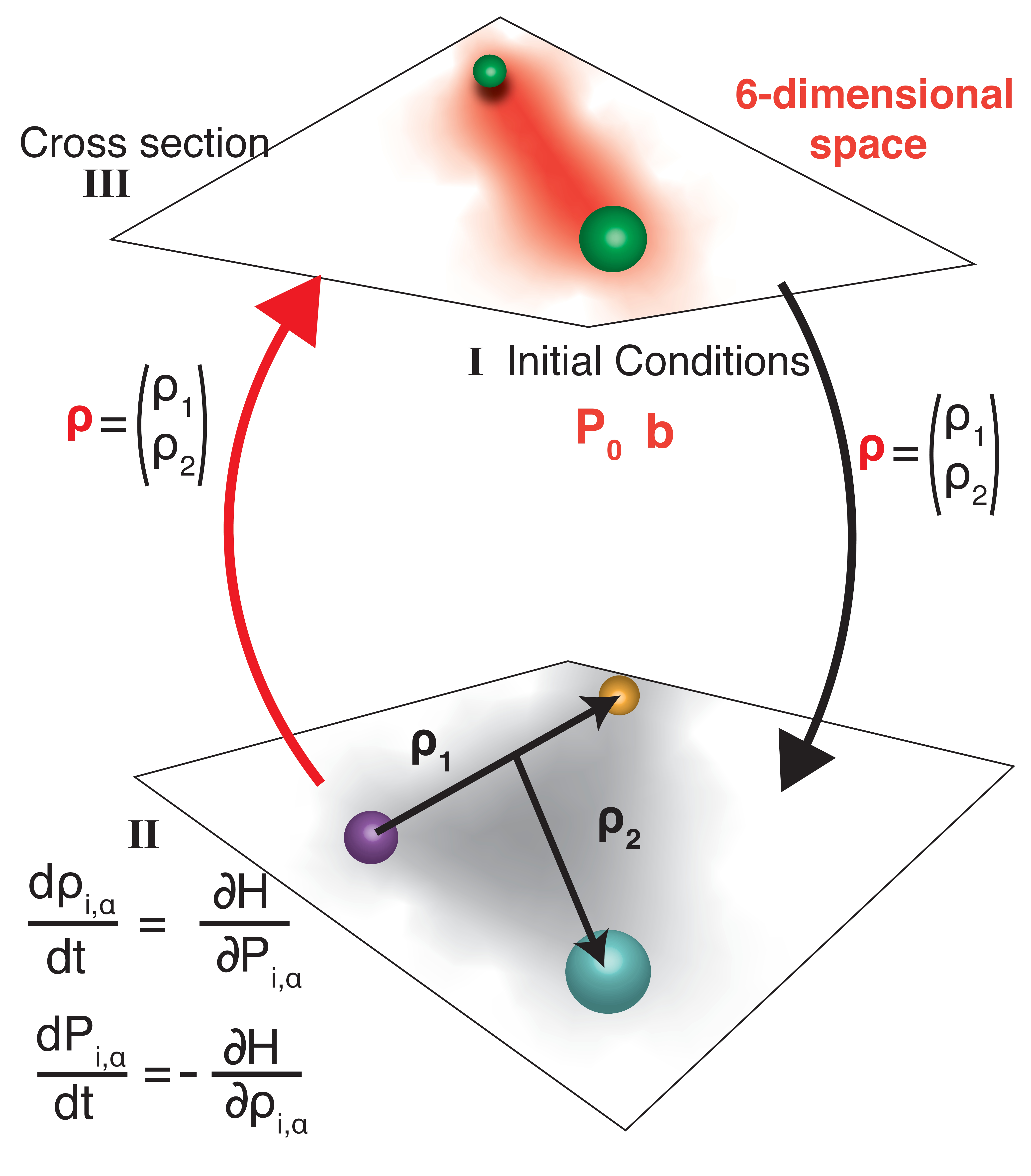}
		\caption{Schematic illustration of the classical trajectory method in hyperspherical coordinates explained in Section~\ref{method}. [With permission, reprinted from \cite{Perez-Rios2020}].}	
		\label{fig:method}	
	\end{center}
\end{figure}

A schematic representation of the method has been illustrated in Fig.~\ref{fig:method}. 
As shown in this figure, in the first step one should determine the initial momentum vector $\vec{P}_0$ and the impact parameter vector $\vec{b}$ of the scattering problem in a 6D space. Note that the orientation of these vectors are sampled randomly from probability distribution functions associated with the appropriate angular elements in hyperspherical coordinates (see Ref.~\cite{Perez-Rios2020}). Without loss of generality, and for the sake of simplicity, the initial Jacobi momentum vector $\vec{P}_2$ (indicating the last three components of the momentum vector) is chosen to be parallel to the $z$ axis in 3D space.

\begin{figure}[t]
	\begin{center}
		\includegraphics[width=1\linewidth]{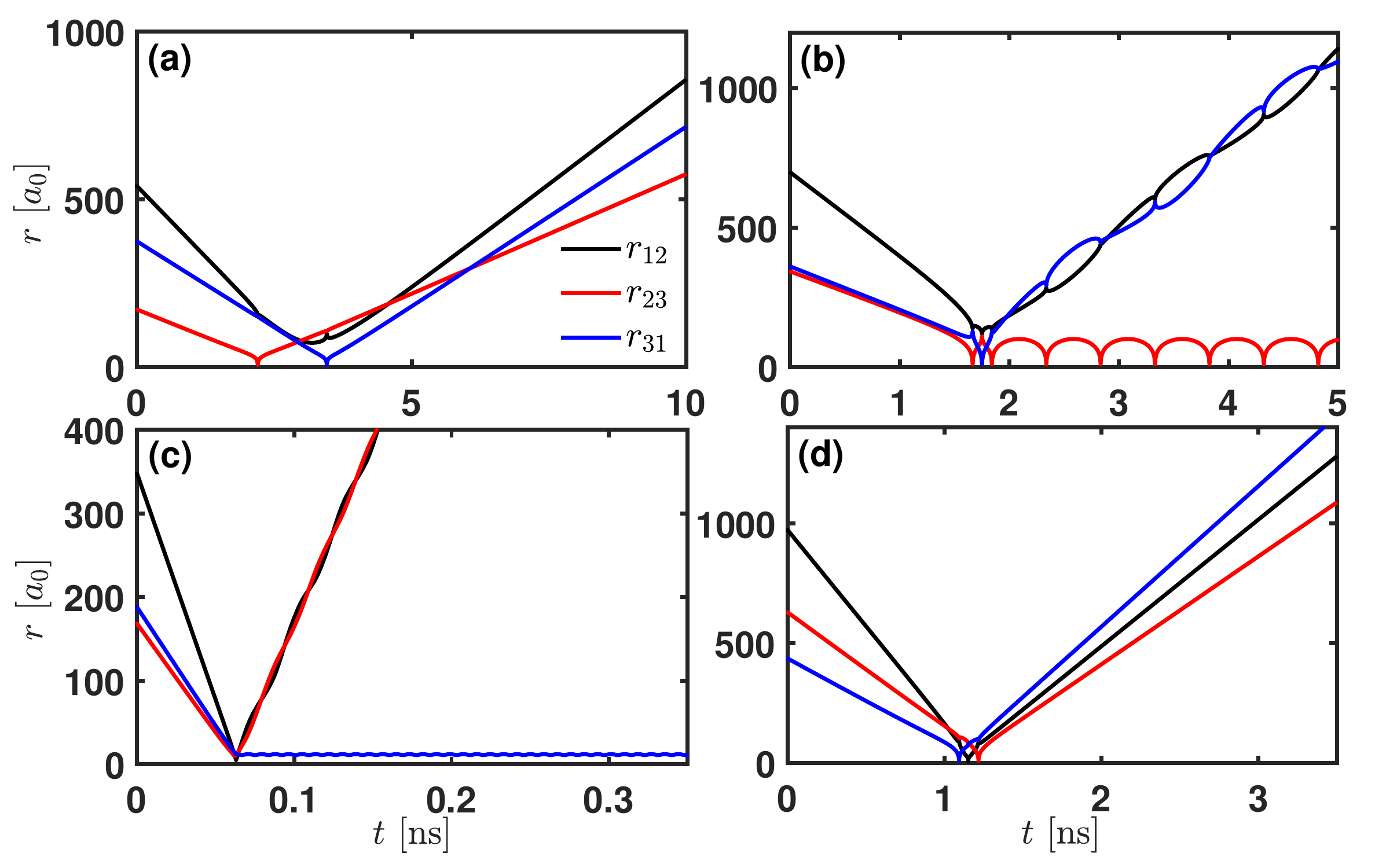}
		\caption{Classical trajectories for (a) P+He+He at $E_c = 0.1$~K with $b=0$, (b) Sr$^+$+Cs+Cs at $E_c = 1$~K with $b = 5~a_0$, (c) Li+He+He at $E_c = 10$~K with $b = 0$, and (d) Yb$^+$+Rb+Rb at $E_c=5$~K with $b=0$. $a_0\approx 5.29\times 10^{-11}$~m is the Bohr radius.}	
		\label{fig:trajs}	
	\end{center}
\end{figure}

Making use of the orthogonality condition $\vec{b}\cdot\vec{P}_0 = 0$ which has been implemented in the calculations, we can obtain the initial position vector in the 6D space from
\begin{equation}\label{eq:position}
	\vec{\rho}_0 = \vec{b} - \dfrac{\sqrt{\rho_0^2-b^2}}{P_0}\vec{P}_0 ~,
\end{equation}
where the length of $\vec{\rho}_0$ is generated randomly from the interval $[R_0-\delta R, R_0+\delta R]$ centered around a suitable $R_0$ which fulfills the condition for three particles to be initially in an uniform rectilinear state of motion. Each point in the 6D space, indicated by $\vec{\rho}_0$ corresponds to a set of all possible configurations of the three particles in the 3D space. The hyperangles specify the shape of the configuration (the triangle made by three particles). The hyperradius describes the size (in a geometrical way) of the triangles, i.e., the enlarging or shrinking of the configuration by the same scale factor in all directions. Therefore, from Eq.~(\ref{eq:rho6D}), we can transform the initial condition to the regular 3D space (following the black arrow), where we solve the Hamilton's equations. In the final step, by using the inverse transformation the solution will be transformed back to the 6D space, where the total cross section and three-body recombination rate are calculated by Eqs.~(\ref{eq:sigma}) and (\ref{eq:k3}), respectively. The opacity function $\mathcal{P}(E_c,b)$ in these equations is calculated as
\begin{align}\label{eq:opacity}
	\mathcal{P}(E_c,b) \approx ~ & \frac{n_r(E_c,b)}{n_t(E_c,b)} \pm \nonumber \\
	& \frac{\sqrt{n_r(E_c,b)}}{n_t(E_c,b)}\sqrt{\frac{n_t(E_c,b)-n_r(E_c,b)}{n_t(E_c,b)}} ~,
\end{align}
where $n_r$ indicate the number of classical trajectories that lead to the recombination events and $n_t$ is the total number of simulated trajectories. 
The second term in Eq.~(\ref{eq:opacity}) is the statistical error due to the stochastic nature of the MC technique~\cite{Perez-Rios2020}. It is important to mention that the total energy and the magnitude of the total angular momentum vector, $J = |\vec{\rho}_1 \times \vec{P}_1 + \vec{\rho}_2 \times \vec{P}_2|$, are conserved during collisions. 

Fig.~\ref{fig:trajs} displays four different examples of trajectories in terms of relative distances $r_{ij} = |\vec{r}_j - \vec{r}_i|$, related to the Jacobi vectors through the following equations  
\begin{align}\label{eq:rel2Jac}
	r_{12} & = \left|\vec{\rho}_1\right| ~, \nonumber \\
	r_{23} & = \left|\vec{\rho}_2- \frac{m_1}{m_1+m_2}\vec{\rho}_1 \right| ~, \nonumber \\
	r_{31} & = \left|\vec{\rho}_2+ \frac{m_2}{m_1+m_2}\vec{\rho}_1 \right| ~.
\end{align}
\noindent
Elastic collisions are depicted in panels (a) and (b). In the first case, there are two consecutive elastic collisions, first between particles $m_2$ and $m_3$ and then between particles $m_3$ and $m_1$. In panel (b), three particles meet nearly simultaneously, and three two-body elastic collisions happen over a brief period. Trajectories leading to three-body recombination are shown in panels (b) and (c). In panel (b), first particles $m_2$ and $m_3$ meet but do not form a bound state, and subsequent $m_1$ and $m_3$ have an elastic collision, and in the final step, $m_2$ and $m_3$ meet again and form a bound state. In panel(c), three particles meet nearly simultaneously, and two of them form a molecule. Note that each product's oscillations' amplitude and frequency differ depending on the reactants and the initial conditions.

\subsection{Grand angular momentum}\label{subsec:GAM}

Scattering in 3D shows an effective potential, including the effect of the centrifugal barrier related to the impact parameters. Similarly, in the case of 6D scattering, the effective three-body potential shows a barrier. However, in this case, it is dominated by the grand angular momentum, initially defined by Smith in Ref.~\cite{Smith1960} in hyperspherical coordinates. Hence, it is necessary to relate the grand angular momentum and impact parameter to develop capture models and threshold laws for the three-body reaction rate. In classical mechanics, the angular momentum in 6D space is a bivector defined by the exterior product (also known as wedge product) of 6D position and momentum vectors as
\begin{equation}\label{eq:grandL}
	\mathrm{\Lambda} = \vec{\rho} \wedge \vec{P}, 
\end{equation} 
which is isomorphic to a $6\times6$ skew-symmetric matrix with elements
\begin{equation}\label{eq:compL}
	\Lambda_{ij} = \rho_i P_j - \rho_j P_i, 
\end{equation}
for $i,j = 1,2,\dots,6$. This general definition applies in all higher-dimensional spaces, and for the 3D space it coincides with the familiar cross product. Note that, $\mathrm{\Lambda}$ is not equal to the ordinary total angular momentum of the three-body system in the 3D space, however, the components of the angular momenta (associated with Jacobi vectors) appear among its elements $\Lambda_{ij}$. Furthermore, by applying the (algebraic) Lagrange's identity~\cite{Gradshteyn2000} one finds
\begin{align}\label{eq:length}
	\mathrm{\Lambda}^2 \equiv \sum_{1\leq i<j\leq6}\Lambda_{ij}^2 
	& = \left(\sum_{i=1}^6 \rho_i^2\right) \left(\sum_{i=1}^6 P_i^2\right)	-\left(\sum_{i=1}^6 \rho_i P_i  \right)^2 \nonumber \\ 
	& = \rho^2P^2 - \left(\vec{\rho}\cdot\vec{P}\right)^2 ~.
\end{align}  

It is worth mentioning that quantum mechanically, the grand angular momentum is an operator and its square, $\mathrm{\Lambda}^2$, is the quadratic Casimir operator of so(6) with hyperspherical harmonics as eigenfunctions and  $\lambda(\lambda+4)$ as eigenvalues ($\lambda$ is a positive integer)~\cite{Dragt1965,Whitten1968,Avery2012}.

\section{Threshold laws for three-body recombination reactions}\label{threshold}

Classical capture models are well-motivated to treat charged-neutral reactions, leading to the well-known Langevin reaction model~\cite{Langevin1905}, predicting an energy-independent three-body recombination rate. A similar approach can be developed for the case of three-body collision. Assuming that the three-body long range interaction is given by $V_{LR} = - C_\mathrm{eff}\rho^\beta$ in hyperspherical coordinates ($\beta<0$), and  including the centrifugal energy, the effective long-range potential can be written as~\cite{Smith1960} (in atomic units) 

	\begin{equation}\label{eq:veff}
		V_\mathrm{eff}(\rho) = V_{LR}(\rho) + \frac{\mathrm{\Lambda}^2}{2\mu\rho^2}	~,
	\end{equation}
	with a maximum (the centrifugal barrier) at $\rho_0 = \left(-\beta\mu C_\mathrm{eff}/\mathrm{\Lambda}^2\right)^{1/(-\beta-2)}$. From Eqs.~(\ref{eq:position}) and (\ref{eq:length}) and conservation of (grand) angular momentum, we find 
	\begin{align}\label{eq:Ebrel}
		\mathrm{\Lambda}^2 & = \left(\vec{b} \wedge \vec{P_0}\right)^2 \nonumber \\
		& = b^2 P_0^2  - \left(\vec{b}\cdot\vec{P_0}\right)^2 \nonumber \\
		& = 2\mu E_c  b^2, 
	\end{align}
	where to write the third line we made use of the orthogonality relation $\vec{b}\perp\vec{P_0}$. In the framework of the classical capture theory, one may define the maximum impact parameter, $b_\mathrm{max}$, as the distance at which the collision energy is comparable to the centrifugal barrier, i.e., $E_c = V_\mathrm{eff}(\rho_0)$. Therefore, we derive the following relation for $b_\mathrm{max}$,  
	\begin{align}\label{eq:bmax}
		b_\mathrm{max} &= \left(\frac{\beta}{\beta+2}\right)^{(\beta+2)/(2\beta)} \left(\frac{-2}{\beta C_\mathrm{eff}}\right)^{1/\beta} E_c^{1/\beta} \nonumber \\
		& = \gamma E_c^{1/\beta}.
	\end{align}
    
    Next, the geometric cross section is obtained by setting $\mathcal{P}(E_c,b) = 1$ for $b\leq b_\mathrm{max}$ (also known as rigid-sphere model) in Eq.~(\ref{eq:sigma}). Upon substituting Eq.~(\ref{eq:bmax}) into Eq.~(\ref{eq:sigma}), we find a threshold law for the three-body recombination cross section, i.e., 
    \begin{align}\label{eq:sigmaLang}
    	\sigma_\mathrm{rec}(E_c)  = \frac{8\pi^2}{3}\int_{0}^{b_\mathrm{max}(E_c)} b^4 d b  = \frac{\gamma^5}{5} E_c^{5/\beta} ~,
    \end{align}
    and, from Eq.~(\ref{eq:k3}), for the energy-dependent three-body recombination rate as  
    \begin{equation}\label{eq:k3Lang}
    	k_3(E_c) = \frac{\gamma^5}{5}\sqrt{\frac{2}{\mu}} E_c^{(10+\beta)/(2\beta)} ~.
    \end{equation}
	Note that the general trend of the three-body recombination rate given in Eq.~(\ref{eq:k3Lang}) is best valid at low collision energies due to the fact that the long-range tail of the potential dominates the recombination rates in this regime, whereas at higher energies the effect of the short-range details of potential on the formation rate of products can not be neglected. This procedure has been applied to develop a classical threshold law for the cases of atom-atom-atom (see Ref.~\cite{Mirahmadi2021}) and ion-atom-atom (see Ref.~\cite{Mirahmadi2023}) three-body recombination. However, still one needs to find the way to find $\beta$, which is explained in the Section below. 
	
	Finally, by plugging Eq.~(\ref{eq:k3Lang}) into Eq.~(\ref{eq:MBLANg}), one finds the temperature dependent three-body recombination rate given by
	
	\begin{equation}\label{eq:k3Tthreshold}
	    k_{3}(T)= \frac{\gamma^5}{10}\sqrt{\frac{2}{\mu}}\Gamma \left( \frac{10+7\beta}{2\beta}\right) (k_BT)^{(10+\beta)/(2\beta)},
	\end{equation}
where $\Gamma(x)$ is the Euler gamma function of argument $x$. Eq.~(\ref{eq:k3Tthreshold}) is general and valid for any three-body potential within the limitations discussed above.

\section{Universality in few-body processes: hyperradial distribution}\label{b3b}
	To find a general expression for the long-range interaction potential $V_{LR}(\rho)$ relevant for the classical trajectory method and the threshold law developed above, one should find the equivalence of the potential $V(\vec{r}_1,\vec{r}_2,\vec{r}_3)$ in the hyperspherical coordinate system. In the majority of cases, potential in the regular 3D space is a function of relative distances $r_{ij}$.  Therefore, we should consider the following equation 
	\begin{equation}\label{eq:vlr_hs}
		V_{LR}(\rho,\alpha_1,\alpha_2,\alpha_3,\alpha_4,\alpha_5) = V(r_{12},r_{23},r_{31}) ~.
	\end{equation}
	Note that, due to the relations given by Eqs.~(\ref{eq:rho6D}) and (\ref{eq:rel2Jac}), the potential in the left hand side of Eq.~(\ref{eq:vlr_hs}) is a function of hyperradius and hyperangles describing the position vector $\vec{\rho}$. However, we are only interested in the hyperradial ($\rho$) dependence of the potential in the 6D space.  
	To tackle this issue, we solve Eq.~(\ref{eq:vlr_hs}) for randomly sampled hyperangles with appropriate weights given by Eq.~(\ref{eq:hsvol}), ensuring a uniform sampling of random points distributed on the 6-sphere (in the geometrical sense). This implies that the solution of Eq.~(\ref{eq:vlr_hs}) will be obtained as a distribution of $\rho$ values.
	
	This method has been adopted to the three- four- and five-body systems interacting via long-range neutral-neutral ($\propto -1/r^6$) and charged-neutral ($\propto -1/r^4$) pairwise potentials, in Refs.~\cite{Mirahmadi2021a,Yu2023,Mirahmadi2023}. Hence, in the 6D space, the hyperradius obtained from this equation is the maximum possible hyperradius for which potential $V_{LR}(\rho)$ affects the particle. Equivalently, in the regular 3D space, it provides the largest configuration (e.g., triangle for three-body case) which beyond that the particles do not interact and can be considered as free particles.
	
\begin{figure}[h]
		\begin{center}
			\includegraphics[width=0.6\linewidth]{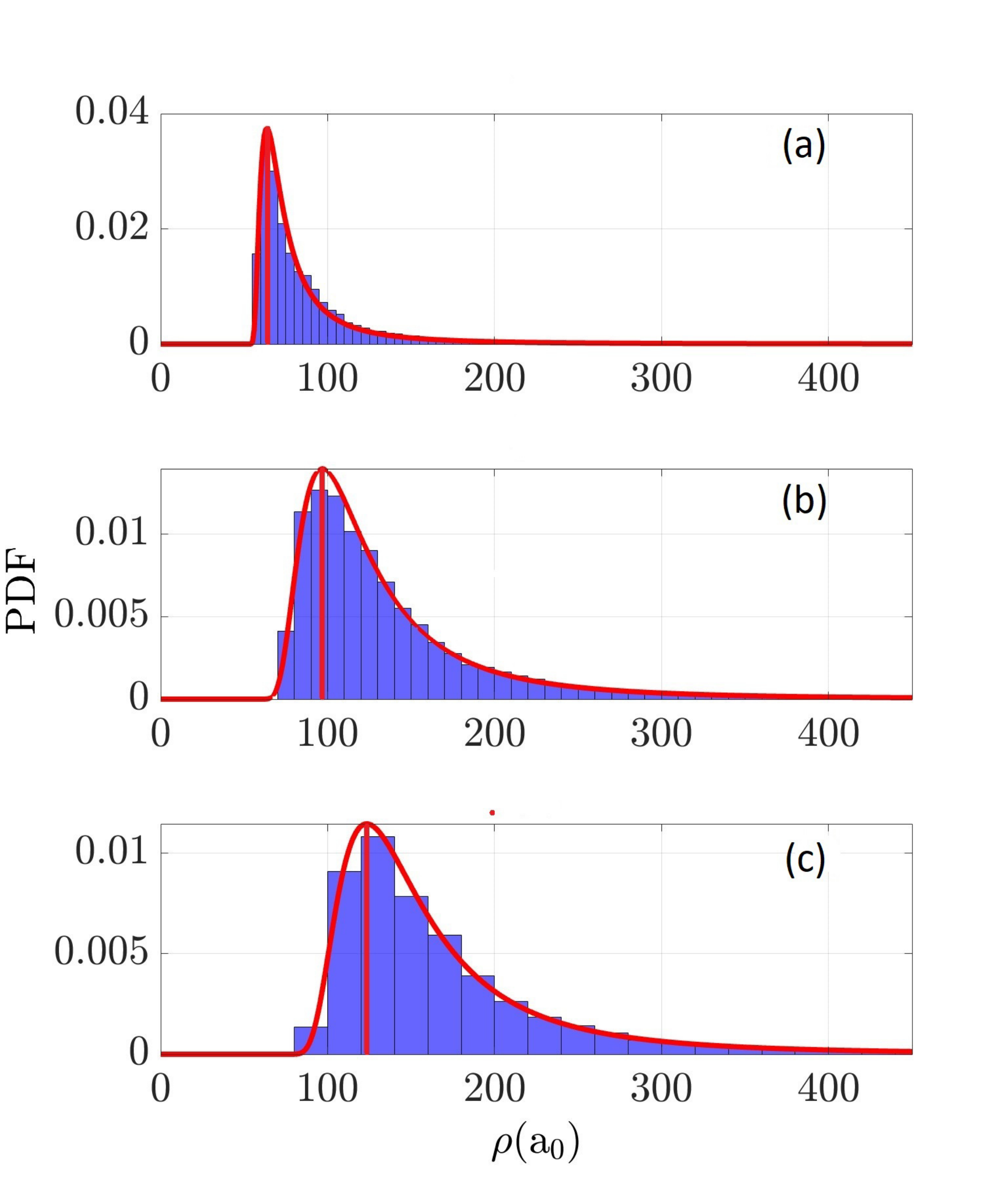}
			\caption{Probability density function (PDF) of the hyperradius distribution for a collision energy of 1~mK for (a) As+He+He, (b) As+He+He+He, and (c) As+He+He+He+He. The red curve represent the fitting to the GEV distribution, whereas the mode is indicated by the vertical lines at 64~a$_0$, 92~a$_0$ and 123~a$_0$ for panels (a), (b) and (c), respectively. [With permission, reprinted from \cite{Yu2023}].}	
			\label{fig:345particles}	
		\end{center}
	\end{figure}

	As a result, we find that, independently of the number and type (charged or neutral) of particles involved, the hyperradial distribution function always follows a special type of the generalized extreme value (GEV) distribution given by
	
	\begin{equation}
\label{eq12}
		f(\rho) = \frac{1}{\delta}\exp\left[-\left(1+\xi\frac{\rho-\beta}{\delta}\right)^{-\frac{1}{\xi}}\right]\left(1+\xi\frac{\rho-\beta}{\delta}\right)^{-1-\frac{1}{\xi}}, 
	\end{equation}
with $1+\xi\frac{\rho-\beta}{\delta}>0$, called the Fr\'echet distribution~\cite{Singh1998}. Hence, the hyperradial distribution function is an universal property of few-body systems independently of the inter-particle interaction and number of particles. Using the Fisher-Tippet-Gnedenko theorem~\cite{Leadbetter2012,LaurensdeHaan2007,Yu2023} and keeping in mind that the effective long-range radius can not be arbitrary small, i.e., the $\rho$-distribution should be bounded from below, one explains why the hyperradial distribution follows a Fr\'echet distribution. As an example, Eq.~(\ref{fig:345particles}) displays the probability density functions of the $\rho$-distribution for three to five-body systems interacting via $1/r^6$ van der Waals forces (for other examples see Ref.~\cite{Yu2023}). The red (vertical) line in three panels (a) to (b) indicate the hyperradius with the maximum likelihood, $\rho_m$, which represents the capture hyperradius. 

\section{Atom-atom-atom three-body recombination reactions}\label{aaa}

Due to cryogenic temperatures and high-density conditions, Buffer gas cells offer a unique arena to study the formation of van der Waals (vdW) molecules via three-body recombination~\cite{DeCarvalho1999,Brahms2008,Brahms2011,Tariq2013,Quiros2017}. These molecules, as the simplest form of vdW complexes, consist of two atoms held together by dispersion interactions (with binding energies $\lesssim 1$ meV) depending on the polarizability of the interacting atoms~\cite{Blaney1976}. Most of these molecules appear when a metal species, X, is ablated in the presence of a He buffer gas, leading to three-body recombination as X + He + He $\rightarrow$ XHe + He.  

\begin{figure}[h]
	 	\begin{center}
	 		\includegraphics[width=0.6\linewidth]{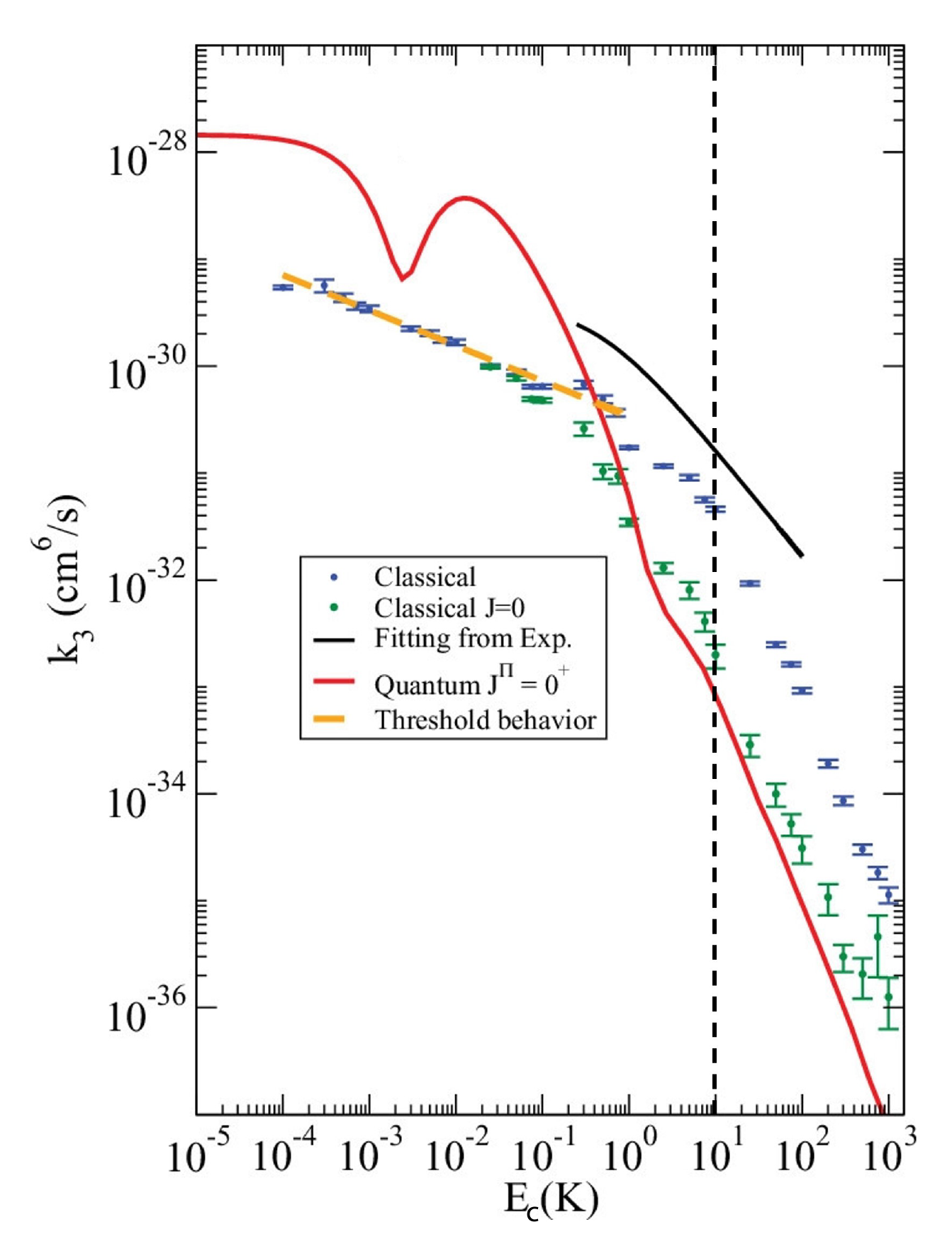}
	 		\caption{\label{fig:3He} Comparison of the three-body recombination rate leading to the formation of He$_2$ molecules through He+He+He reaction, computed using different methods: quantum calculations based on the adiabatic hyperspherical method (solid red line), classical trajectory calculations for direct three-body recombination (blue points), and the same classical trajectory calculations but for a fixed total angular momentum $J=0$ (dark green points). The solid black curve is a fitting of the results produced by a theoretical model from Ref.~\cite{Bruch2002} based on the simulations with inputs from experimental values as well as free model-dependent parameters.  Black vertical dashed line indicates the dissociation energy of the He$_2$, i.e., $D_\mathrm{He_2}\approx 10.49$~K [Reproduced partially with permission from Ref.~\cite{Perez-Rios2014}.]}
	 	\end{center}
	 \end{figure}

Assuming a pairwise potential $V(\vec{r}_1,\vec{r}_2,\vec{r}_3)$ in Eq.~(\ref{eq:cartesianH}) as 
	 \begin{equation}\label{eq:add_pot}
	 	V(\vec{r}_1,\vec{r}_2,\vec{r}_3) = U(r_{12}) +  U(r_{23}) +  U(r_{31}) ~,
	 \end{equation}
where $U(r_{ij})$ stand for the two-body potentials, it is found that the energy-dependent formation rate, $k_3(E_c)$, of XHe molecules, shows two different power-law behaviors associated with two energy regimes: the low energy $E_c<D_{\text{XHe}}$ and the high energy $E_c>D_{\text{XHe}}$ regime, where $D_{\text{XHe}}$ represents the dissociation energy of the XHe molecule. This behavior is shown in Fig.~\ref{fig:3He} for the three-body recombination of He atoms, leading to the formation of He$_2$, a prototypical vdW molecule. We notice that the classical calculations can reproduce the same trend as the quantum calculations for $E_c\gtrsim 10$~K, corresponding with the dissociation energy of He$_2$ molecule~\cite{Aziz1995} and depicted as the vertical dashed line. However, the values of classical and quantum rates do not show a quantitative agreement. This can be improved if the classical calculations are done for a fixed total angular momentum $J=0$, mimicking the angular momentum and parity quantum number for the quantal calculations (dark green data set in Fig.~\ref{fig:3He}). On the contrary, the quantum and classical calculations do not agree in the low energy regime, neither qualitatively nor quantitatively, as one expected. Indeed, the discrepancy between classical and quantum calculations appears at collision energies of the order of 1~K, which corresponds with the van der Waals energy for He-He collisions~\cite{Perez-Rios2014}.

Fig.~\ref{fig:3He} shows that the formation rate of molecules is relatively insensitive to short-range interactions and is dominated by the long-range tail of the potentials ($-1/r^6$). Whereas, at higher collision energies, the short-range part of the potential plays an important role. The classical threshold law derived in Section~\ref{threshold} predicts a three-body recombination rate $\propto E^{-1/3}$, depicted by the orange dashed line and very well describes the trend of the energy-dependent recombination rate at the low energy regime. It is worth mentioning that the same energy dependence is obtained when only the two-body potential is considered to extract the maximum impact parameter~\cite{Mirahmadi2021}. However, in this case, the rate is quantitatively less accurate than the one presented in Section~\ref{threshold}.

\begin{figure}[t]
	 	\begin{center}
	 		\includegraphics[scale=0.5]{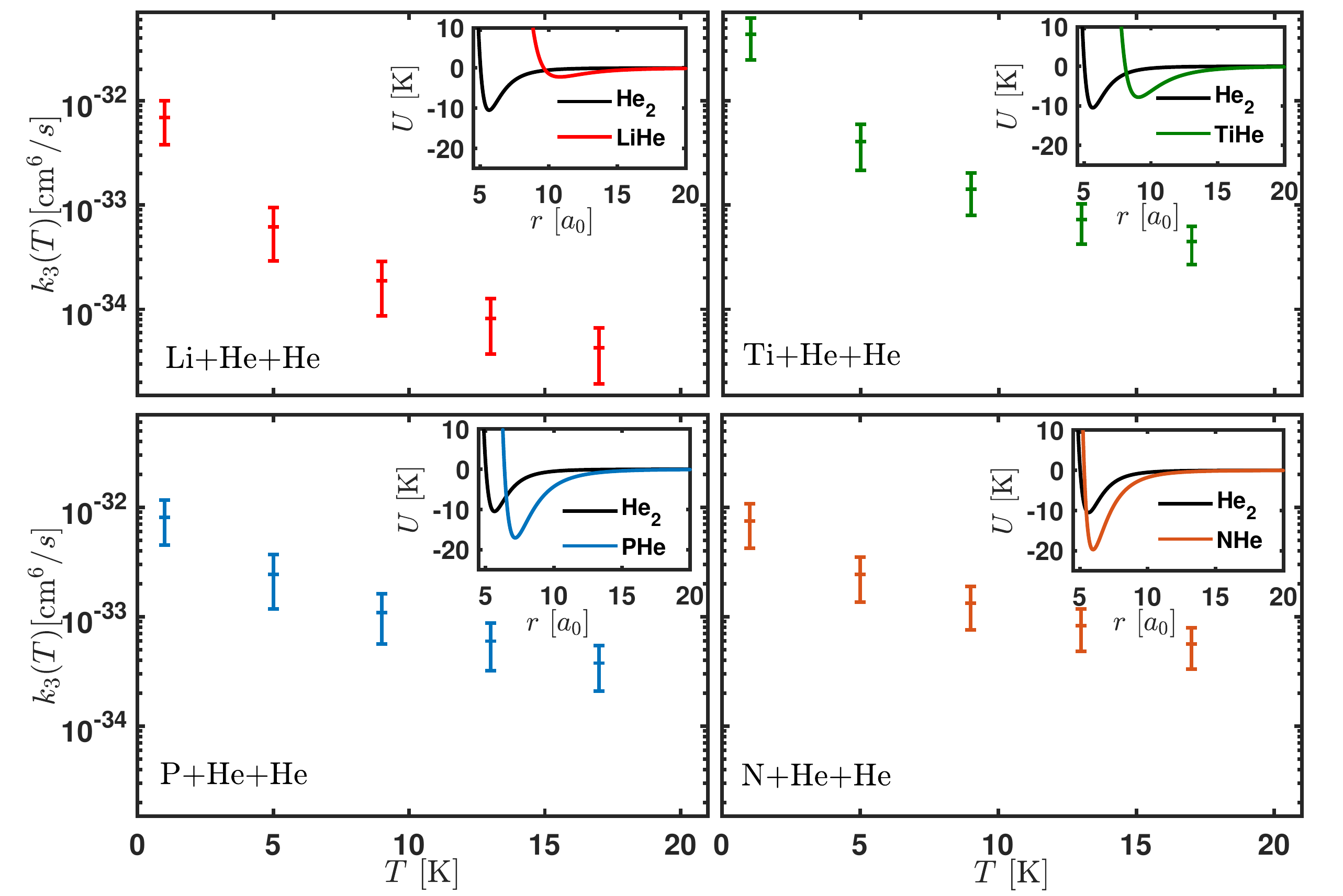}
	 		\caption{\label{fig:4atomk3T} Three-body recombination rates of four different XHe van der Waals molecules as a function of temperatures $T\in\{4,8,12,16,20\}$~K (semi-log plots). Potentials for He-He and He-X are taken from \cite{Mirahmadi2021}. [With permission, reproduced from \cite{Mirahmadi2021}.]}
	 	\end{center}
	 \end{figure}

A more general study has been carried out considering X + He + He three-body processes, where X consists of alkali, alkaline-earth, transition metals, pnictogens, chalcogens and halogen atoms. The results confirmed that for temperatures relevant to the buffer gas cells (1K~$\lesssim T \lesssim 10$ K), almost any vdW molecule, XHe, is produced with the same rate via three-body recombination, as shown in Fig.~\ref{fig:4atomk3T}. This figure displays the temperature-dependent rates [see Eq.~(\ref{eq:MBLANg}] $k_3(T)$ for several XHe examples: LiHe, TiHe, PHe and NHe. Similar results have been reported for different vdW molecules in the limit of zero collision energy using a full quantum treatment~\cite{Bin2022}. However, some deviations are observed for HeBa and HeBe. 

LiHe,~\cite{Tariq2013} AgHe,~\cite{Brahms2008} and TiHe,~\cite{Quiros2017} molecules have been observed in buffer gas cell experiments after ablation of a metal target (Li, Ag, and Ti) in the presence of a cryogenic He buffer gas. Hence, based on our results for the three-body recombination of X and He atoms, ablating any metal X in the buffer gas cell will potentially lead to detectable XHe molecules. In particular, they are detectable because they will be produced sufficiently to be traceable through spectroscopic methods. First, however, it is necessary to correctly identify the molecular dissociation processes to understand the vdW molecules in equilibrium conditions. In other words, the molecules' observation is contingent on their bending energy and the temperature of the buffer gas cell. Only at low temperatures, extremely weakly bound molecules could be observed. Otherwise, despite the formation of XHe molecules, these will dissociate through collisions with He atoms.

\section{Charged-neutral-neutral three-body recombination}\label{cnn}

\subsection{Cold Chemistry}\label{coldchemistry}
Thanks to the development of hybrid atom-ion traps, it is possible to study charged-neutral reactions down to the millikelvin regime, giving rise to the so-called cold chemistry. That is the study of chemical reactions between 1~K and 1~mK. Indeed, it is possible to go as low as micro-kelvin energies when the ion is very heavy in comparison with the mass of the atoms~\cite{Fedker2020,Weckesser2021}. Moreover, at these low temperatures, the average kinetic energy of atoms and molecules is comparable with the typical energy shifts caused by external fields. As a result, it is possible to study of charged-neutral interactions in a controllable manner, leading to the discovery of novel reaction mechanisms~\cite{Cote2000,Stefan2008,RevModPhysatomion,COTE20166,Sikorsky2018,Gacesa2016,Sourav2016,Kleinbach2018,Puri1370,Stefan2019,Hirzler2020,Hirzler2021,Saito2017}.

\begin{figure}[t]
\begin{center}
 \includegraphics[width=0.9\linewidth]{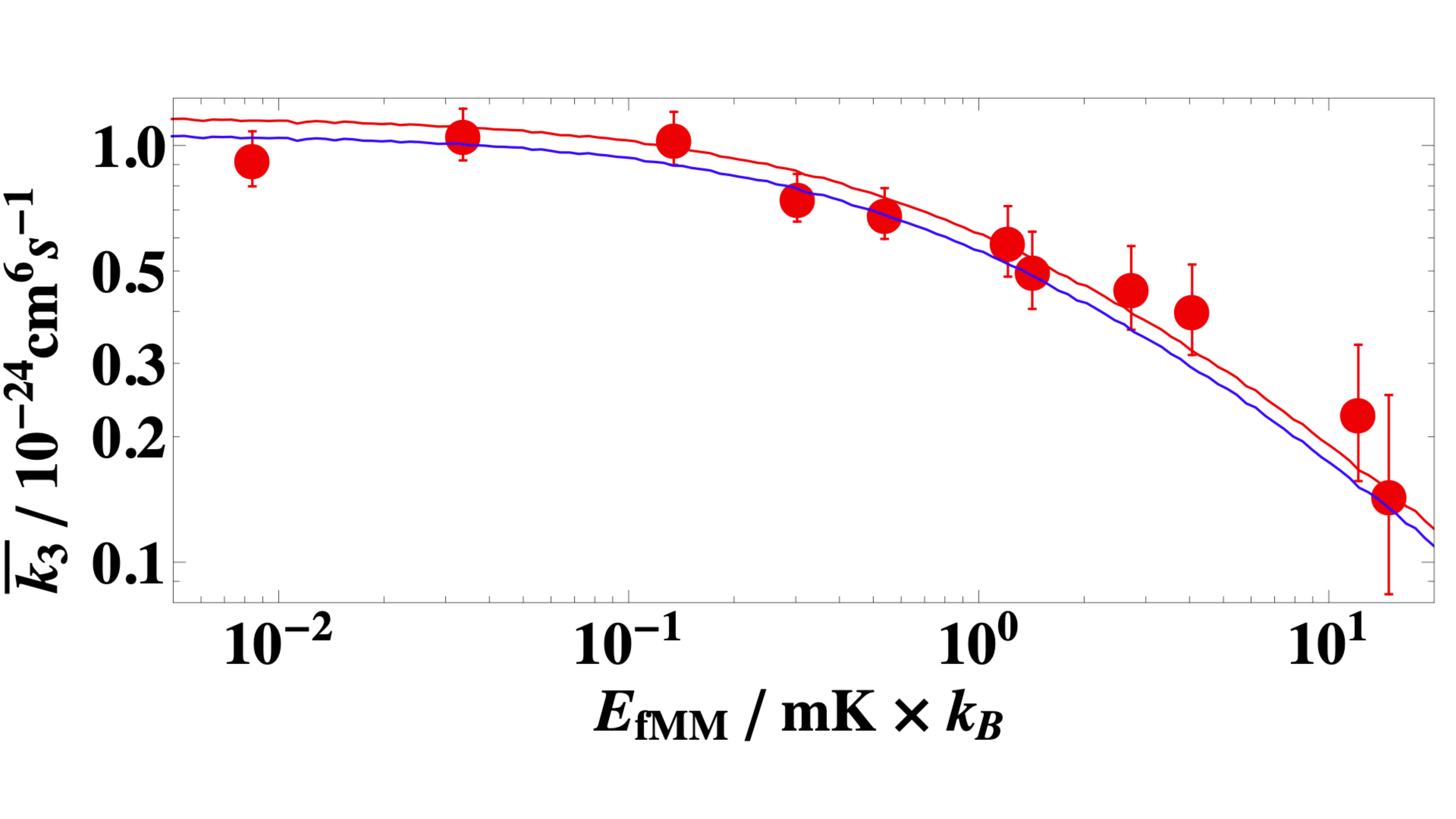}
\caption{\label{fig9} Three-body recombination rate for the reaction Ba$^+$ + Rb + Rb $\rightarrow$ BaRb$^+$ + Rb, as a function of the excess of micromotion energy of the ion in the Paul trap. Experimental measurements are shown as the red points and the red line represents a fitting to the points. On the contrary, a first principle prediction based on classical trajectory calculations in hyperspherical coordinates of appears as the blue line. Figure reproduced with permission from Ref.~\cite{Krukow2016} }
\end{center}
\end{figure}

Among many groups and authors devoted to the study of cold chemistry involving charged and neutrals, it is worth highlighting the work of Prof. Johannes Hecker Denschlag's group, in which a single Ba$^+$ was brought in contact with an ultracold and highly dense cloud of Rb atoms. The particular interest of this experiment is that charge transfer reactions, Ba$^+$ + Rb$\rightarrow $ Ba +Rb$^+$, are highly suppressed since it has to be photon-mediated. As a result, the Ba$^+$ ion can live longer in the sea of Rb atoms. However, the ion was lost after a certain interaction time despite its negligible two-body loss. In turned out to be a cause of ion-atom-atom three-body recombination, Ba$^+$ + Rb + Rb $\rightarrow$ BaRb$^+$ + Rb, as shown in Ref.~\cite{Krukow2016}, and summarize in Fig.~\ref{fig9}. This figure compares the theoretical direct three-body recombination results based on a pair-wise approximation for the three-body potential, as shown in Eq.~(\ref{eq:add_pot}), and the experimentally derived three-body recombination rate, leading to an outstanding agreement between a first principle approach and experimental results.

The agreement between the classical trajectory method and the experimental results can be rationalized by exploring the validity of the classical approach for the scenario at hand. In particular, it is possible to address the reliability of a classical approach by looking into the number of contributing partial waves to the scattering cross section. Indeed, many partial waves dominate the ion-atom dynamics at low temperatures; thus, making a classical approach viable for reactions in cold environments~\cite{Perez-Rios2020,Perez-Rios2021,JPR2019}.

Theoretical investigations based on a direct approach for ion-atom-atom three-body recombination predicts molecular ions as the main reaction product~\cite{Perez-Rios2015,Perez-Rios2018,Mirahmadi2023}. A threshold law based on the dominance of charged-induced dipole interaction versus induce dipole-induce dipole interaction --a simplified version of the more general approach presented in Section~\ref{threshold}, was derived, leading to the same conclusion: the main reaction product in ion-atom-atom three-body recombination~\cite{Perez-Rios2015,Perez-Rios2018}. However, experiments did not show any trace of molecular ions. The situation changed when experiments were conducted in the dark, i.e., in the absence of light, creating the dipole trap for the atoms, leading to the observation of molecular ions. Therefore, it showed that the evolution of a single ion in a high density media is more complex than initially thought. To fully understand the evolution of a single ion, it is necessary to include three-body recombination~\cite{Krukow2016,Perez-Rios2015}, vibrational quenching and dissociation of the molecular ion via collisions with neutrals~\cite{JPR2019}, spin-exchange collisions~\cite{Mohammadi2021}, photo-dissociation of the molecular ion due to the trapping lasers for atoms and cooling ones for the ion~\cite{Mohammadi2021,daSilva2017}, and even the role of the Paul trap as a possible source of molecular ion dissociation~\cite{JPR2021}.

Ion-atom-atom three-body recombination shows to reaction products: B$^+$ + A + A $\rightarrow$ AB$^+$ + A and B$^+$ + A + A $\rightarrow$ A$_2$ + B$^+$. However, classical trajectory calculations predict that mainly AB$^+$ appear as the reaction product. The same result appears invoking a capture model leading to a threshold law as the one derived in Section~\ref{threshold}. Indeed, as shown above, the preponderance of molecular ions over molecules as reaction products has been confirmed experimentally~\cite{Mohammadi2021}. However, the theoretical predictions and experimental observations do not explain how and why molecular ions dominate the ion-atom-atom three-body recombination. Recently, we have conducted systematic calculations on ion-atom-atom three-body recombination reactions in a wide range of ion-atom and atom-atom interaction potentials, finding an explanation of such behavior.

	\begin{figure}[h]
		\begin{center}
			\includegraphics[width=0.75\linewidth]{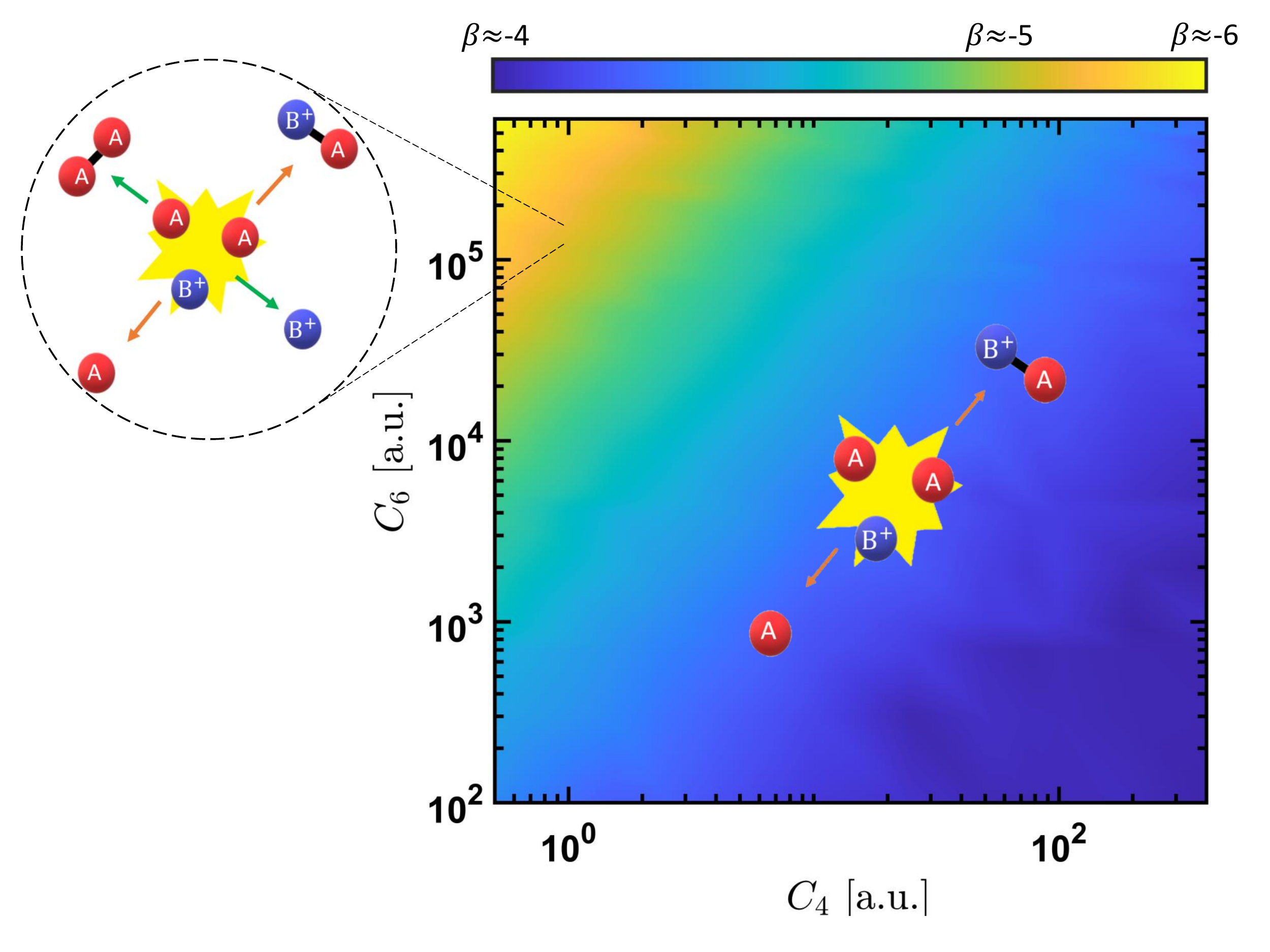}
			\caption{\label{fig:Pdiag} Heat map of visualizing the parameter $\beta$ as a function of long-range two-body interaction coefficients $C_4$ and $C_6$ in the log-log scale. The schematic illustrations display the dominant reactions at low collision energies. [With permission, reproduced from \cite{Mirahmadi2023}.] }
		\end{center}
	\end{figure}

Assuming $V(\vec{r}_1,,\vec{r}_2,\vec{r}_3)$ as the sum of the pairwise potentials, with the long-range tails $-C_6/r^6$ and $-C_4/r^4$ for the A-A and A-B$^+$ interactions, respectively, we obtain the hyperradial potential $V_{LR}(\rho)\propto \rho^\beta$, in which $\beta $ is a function $\beta(C_6,C_4)$. Fig.~\ref{fig:Pdiag} displays this parameter over the intervals $C_6\in [10^2,6\times10^5]$ and  $C_4\in[0.5,400]$, in atomic units ($C_6=C_6^\mathrm{A_2}$ is the van der Waals dispersion coefficient and $C_4=C_4^\mathrm{AB^+}$ is half of the atom (A) polarizability). Three main regimes are noticeable in this Figure: (I) $\beta\approx-4$ (blue color) represents a typical scenario in which the charged-neutral interaction dominates the course of the reaction, leading mainly to the formation of ions, as sketched in Fig.~\ref{fig:Pdiag}, (II) $\beta\approx-6$ (yellow color) means that the neutral-neutral interaction is the most significant interaction, which translates into a larger production of neutral molecules, and (III) $\beta\approx -5$ (greenish yellow color), i.e., an intermediate regime, when both neutral-neutral and neutral-charged interaction have a considerable contribution, the three-body recombination should lead to a similar amount of neutral molecules to molecular ions. 

For $\beta=-4$, Eqs.~(\ref{eq:sigmaLang}) and (\ref{eq:k3Lang}), read as $\sigma_\mathrm{rec}(E_c)\propto E_c^{-5/4}$ and $k_3(E_c) \propto E_c^{-3/4}$, respectively, verifying the threshold law given in Refs.~\cite{Perez-Rios2015,Perez-Rios2018}, under the assumption that only ion-atom interaction dictates the outcome of the three-body recombination. Note that the rate provided by Eq.~(\ref{eq:k3Lang}) accounts for both A$_2$ and AB$^+$ products of the three-body recombination. However, in this scenario, molecular ions, AB$^+$, are the main reaction products, and the formation of $A_2$ molecules can be neglected. Generally, the coefficients $C_6$ and $C_4$ in most ion-atom-atom reactions are associated with $\beta\approx-4$, thus explaining why ion-atom-atom three-body recombination leads to the formation of molecular ions. 

Finally, it is worth mentioning that the preponderance of molecular ions versus molecules has been predicted in the case of ultracold He + He + Li/H$^-$ three-body recombination using an adiabatic hyperspherical full quantal treatment~\cite{Bin2019,Bin2021}, but in the case of He + He + D$^-$, the formation of He$_2$ molecule is slightly larger than the molecular anion~\cite{Bin2017}.

\subsection{Hyper-thermal chemistry}\label{hyper}

Surprisingly enough, the derived threshold law, $\sigma_\mathrm{rec}(E_c)\propto E_c^{-5/4}$ and $k_3(E_c) \propto E_c^{-3/4}$, verifies in room temperature experiments~\cite{Perez-Rios2018}, like in ion mobility experiments of ions in their parent gas. Thus, this implies that the transition between long-range dominated toward short-range dominated physics occurs at higher energies than in the case of neutral-neutral collisions. To understand this intriguing fact, we have explored ion-atom-atom three-body recombination in a wide range of collision energies ranging from the cold to the hyper-thermal regime ($E_c \approx 10^4$~K).

    	\begin{figure}[h]
    	\begin{center}
    		\includegraphics[scale=0.6]{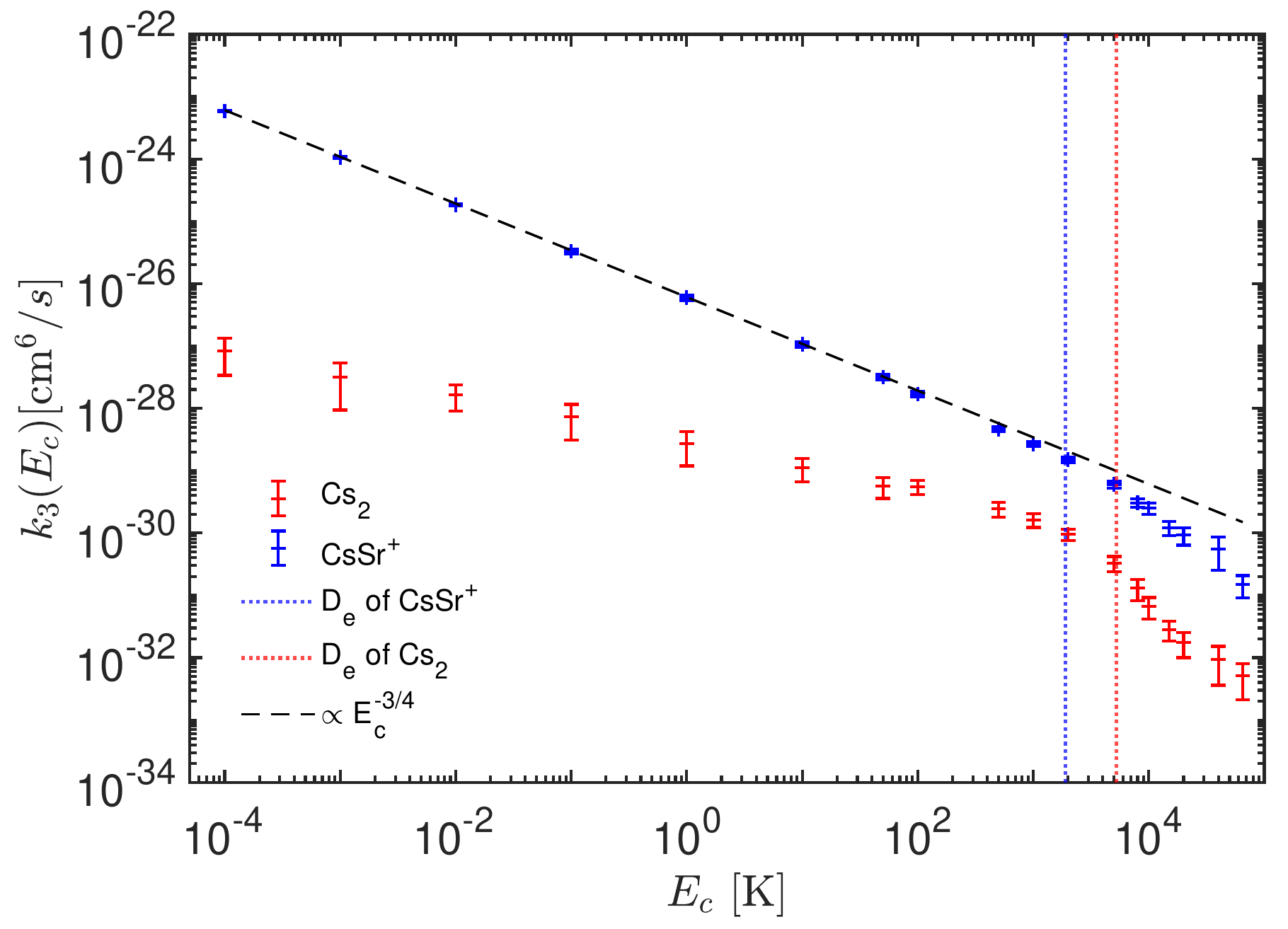}
    		\caption{\label{fig:CsSr} Three-body recombination rates $k_3(E_c)$ for the Cs+Cs+Sr$^+$ reaction. The details for the Cs$_2$ and CsSr$^+$ potentials are given in \cite{Mirahmadi2023}. Error-bars are associated with the Monte Carlo nature of the approach shown in Eq.~(\ref{eq:opacity}). The black dashed line indicates the power-law given in Eq.~(\ref{eq:k3Lang}). The blue and red vertical dashed lines indicate the dissociation energies of CsSr$^+$ and Cs$_2$, respectively. [With permission, reprinted from \cite{Mirahmadi2023}.]  }
    	\end{center}
    \end{figure}
   
   As an example, Fig.~\ref{fig:CsSr} displays the formation rates of CsSr$^+$ (indicated by blue color) and Cs$_2$ (red color) molecules through the direct three-body recombination Cs + Cs + Sr$^+$. Looking into CsSr$^+$ three-body recombination rate, like the case of atom-atom-atom three-body collisions, one can identify two regimes associated with two different power-law behaviors (linear in the log-log scale) distinguished by the dissociation energy of the CsSr$^+$ potential. Similarly, the two energy regimes can be recognized through the three-body recombination rates of Cs$_2$, with a different trend than those of the CsSr$^+$ formation. In particular, the behavior of $k_3(E_c)$ for the formation of neutral molecules changes twice: one slight change near the dissociation energy of molecular ion and a pronounced change around the dissociation energy of the neutral molecule itself.   
   
  On the other hand, it has been shown that the dissociation energy of the molecular ion establishes the threshold energy between the low-energy regime and high-energy regime~\cite{Mirahmadi2023}. That is the threshold energy between long-range and short-range dominated physics. Hence, explaining why the threshold law is still fulfilled in noble gas ions in their parent gases at 300~K~\cite{Perez-Rios2015,Perez-Rios2018} since the dissociation energy of the molecular ion is much larger than the maximum collision energy for those experiments. These investigations confirm that the formation rate of molecular ions through the direct three-body process A + A + B$^+$ in the low-energy regime is dominated by the long-range tail of the potentials. It shows the general trend ($\propto E_c^{-3/4}$ with $\beta=-4$), independent of the A and B$^+$ species under consideration. However, this is not true when the systems under consideration enter a new regime at high collision energies dominated by short-range physics. Hence, it is necessary to consider both reactions B$^+$ + A + A $\rightarrow$ AB$^+$ + A and  B$^{+}$+ A + A$^+$ $\rightarrow$ A$_2$ + B$^+$ in the hyper-thermal region since molecular ions and molecule are produced with nearly the same rate.  

\subsection{Plasma physics}\label{plasma}

In plasmas, where electrons, neutrals, and ions live in extreme conditions of density and temperature, it is well-known that three-body recombination appears in the form of the so-called volumetric recombination, i.e.,  e + H$^+$ + e $\rightarrow$ H + e, one of the primary mechanisms o plasma neutralization~\cite{Krasheninnikov1997,krasheninnikov2017,KUKUSHKIN2017984}. However, in some particular circumstances, other three-body processes may emerge. For instance, in tokamaks, a divertor controls the heat transfer from the plasma to the reactor's walls, avoiding the pollution of the core plasma. In the divertor, ions impinging on the surface of the divertor neutralize, i.e., H$^+$ $\rightarrow$ H (surface mediated), giving rise to possible H + H$^+$ + H $\rightarrow$ H$_2^+$ + H reactions~\cite{Krsti2003,Cretu2022}. These reactions could potentially impact the momentum transfer of plasma at the divertor and with it, the probability of sustaining the reaction. 

\begin{figure}[t]
\begin{center}
 \includegraphics[width=0.8\linewidth]{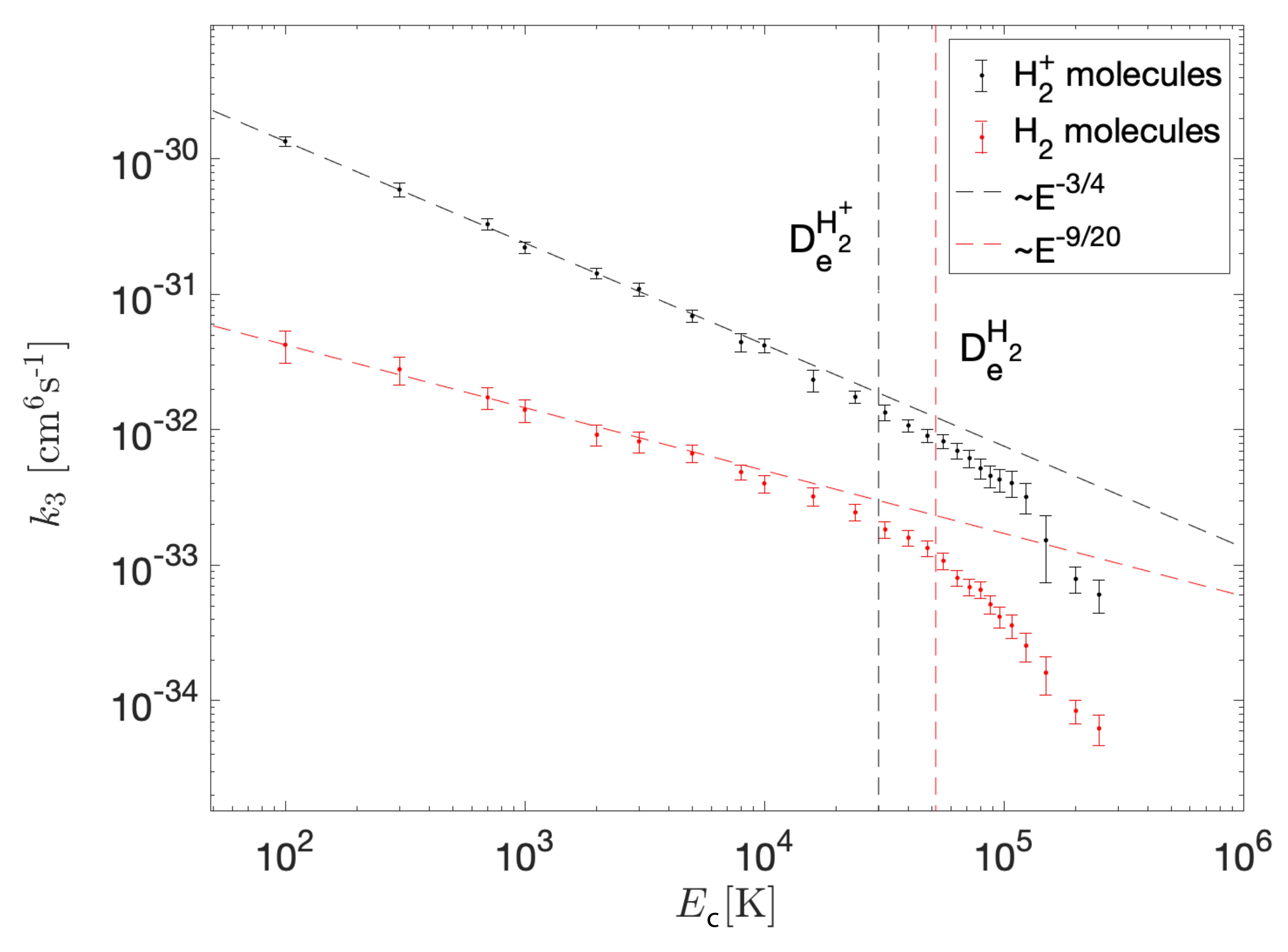}
\caption{\label{figplasma} Energy-dependent ion-atom-atom three-body recombination rates for the hydrogen plasmas. The black symbols depict the rate of molecular ion formation, whereas the red symbols describe the formation of molecules. The error bars attached to each symbol account for one standard deviation error as customary in Monte Carlo simulations. The vertical lines in the left panel represent the dissociation energies of H$_2^+$ and H$_2$. The dashed lines represent two distinct threshold behavior regarding the dominant two-body interaction. Figure reproduced from Ref.~\cite{Cretu2022} }
\end{center}
\end{figure}

Using a direct approach for three-body processes, we have explored three-body recombination in cold hydrogen and deuterium plasmas. The results for hydrogen plasmas are shown in Fig.~\ref{figplasma}. First, as explained above, we notice a change in the trend of the energy-dependent three-body recombination rate, $k_3(E_c)$, around the dissociation energy of the molecular ion. For collision energies less than the dissociation energy of the molecular ion, we notice that  $k_3(E_c)\propto E_c^{-3/4}$, as one expects from the threshold law explained above. However, for molecular formation, we notice that $k_3(E_c)\propto E_c^{-9/20}$, being very similar to the expected threshold law for an interaction potential $V(r)\propto r^{-5}$, although any of the potentials included show this behavior. On the contrary, once the collision energy surpasses the dissociation energy of the molecular ion product, a new regime appears dominated by the short-range details of the potential. In this new regime, the reaction products are in the same amount, more or less, molecular ions and molecules. It is worth mentioning that the same has been obtained for the case of cold deuterium plasmas~\cite{Cretu2022}.

The temperature-dependent rate has been calculated showing a temperature dependence $\propto T^{-3/4}$~\cite{Cretu2022} in stark contrast to volumetric plasma recombination processes $T^{-9/2}$~\cite{Sayasov}. However, the rate of the former is much lower than the latter. Therefore, at high plasma temperatures, ion-atom-atom three-body recombination may play a relevant role in plasma dynamics.

\section{Three-body recombination including internal degrees of freedom}\label{degreesoffreedom}

Three-body recombination processes go beyond atomic recombination or ion-atom-atom processes, including colliding partners with internal degrees of freedom. In particular, molecule-atom-atom, molecule-molecule-atom, and molecule-molecule-molecule three-body recombination have been experimentally explored and, in some cases, theoretically. This section summarizes the primary experimental and theoretical efforts toward understanding termolecular reactions in which the colliding partners have internal degrees of freedom. 

\subsection{Molecule-atom-atom three-body recombination}\label{maa}

\subsubsection{Ozone formation}\label{ozone}
    Ozone molecule (O$_3$) plays a crucial role in atmospheric physics due to its significant impact on the atmosphere's chemistry and climate change. For instance, ozone is essential in the stratosphere to absorb the damaging UV radiation before reaching the earth's surface\cite{BAR13:172}. On the contrary, ozone is a greenhouse gas and air pollutant in the troposphere~\cite{barnes2019ozone}. Therefore, it is crucial to understand how ozone is formed to fully understand its impact in life on earth. 
    
    It is known that ozone is formed through the three-body recombination O$_2$ + O + M $\rightarrow $O$_3$ + M, where M is typically N$_2$ or O$_2$. However, despite its importance and decades of research, the reaction mechanism behind ozone formation is not fully understood due to its complexity. Ozone formation reaction has been studied from an indirect approach, leading to two distinct reaction mechanisms: the energy-transfer mechanism (also known as the Lindemann or the stabilization mechanism) and the Chaperon mechanism, also known as the radical-complex mechanism (see, e.g., Refs.~\cite{MAU05:1,LUT05:2764,TEP16:19194,hippler1990temperature}). According to the energy-transfer mechanism, a metaestable ozone molecule O$_3^*$ is formed at the first step in Eq.~(\ref{eq:2step}), which is stabilized at the second step by a collision with atom M. On the other hand, in the Chaperon mechanism, there are two possible intermediate complexes: O$_2$M$^*$ or OM$^*$. Then, each of these intermediate complexes experience a second or even a third collision to finally give rise to O$_3$. The relevance of the energy-transfer versus the Chaperon mechanism at given conditions relies on the partial pressures and temperature. However, it is not straightforward to distinguish the importance of these two mechanisms by conducting experiments. That is, it is necessary to make some theoretical assumptions, or to develop a first-principle explanation of the ozone formation reaction.

     \begin{figure}[h]
    	\centering
    	\includegraphics[scale=0.5]{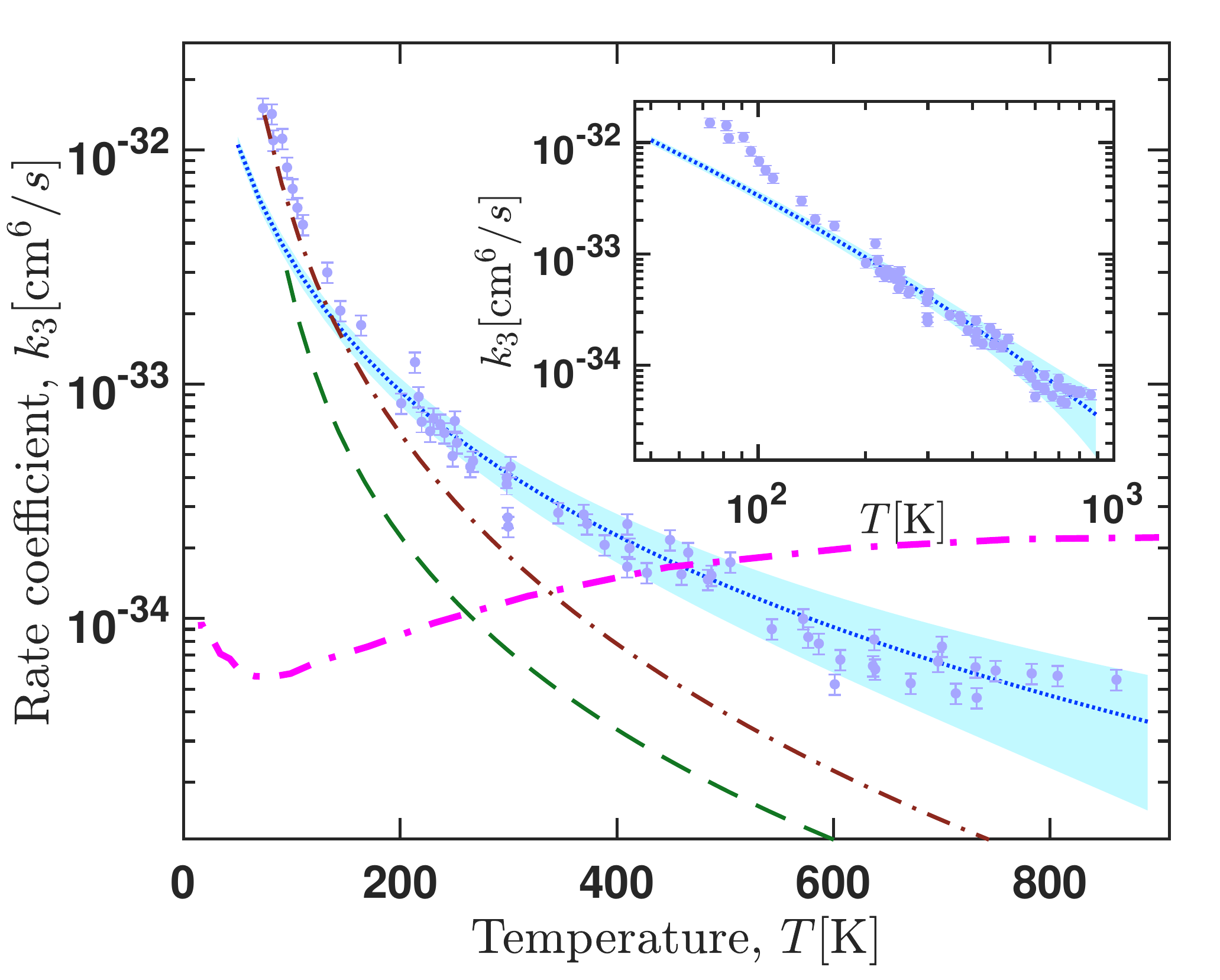}
    	\caption{Recombination rate coefficients for ozone formation. The shaded area (light blue) represents the uncertainty of the theoretical model. The circles with error bars are the experimental data taken from several references given in Ref.~\cite{Mirahmadi2022}. Previous theoretical results are shown with a thick dash-dotted (magenta) line \cite{CHA04:2700}, thin dash-dotted (brown) line \cite{LUT05:2764}, and dashed (green) line \cite{ivanov2006}. The inset shows the data in the log-log scale. [With permission, reprinted with few modifications from \cite{Mirahmadi2022}.] }
    	\label{fig:k3T_QU}
    \end{figure}  
    
Recently, the ozone formation reaction has been explored from a direct approach, enlightening the relevance of the energy-transfer and Chaperon mechanism as a function of the temperature~\cite{Mirahmadi2022}. The theoretical approach combines an {\it ab initio} potential energy surface of the ArO$_3$ system and the classical-trajectory method in the hyperspherical coordinates explained above. In this case, the O$_2$ molecule is treated as a super-atom reducing the degrees of freedom of the O+O$_2$+Ar system. This results in a fixed angle between O$_2$ and the direction to O, corresponding to the bond angle in O$_3$ (117$^\circ$). Therefore, we compute the three-body recombination rate for different angles in the PES and then average over it. Thus, the role of the internal degrees of freedom of O$_2$ is only accounted as an effective interaction rather than explicitly, which is a valid argument for the temperature range used here (vibrational excitation of O$_2$ requires energies $\gtrsim$~1400~K). The potential energy surface used to produce these results is constructed as the sum $V=V_\mathrm{O-O_2}+V_\mathrm{Ar-O_2}+V_\mathrm{Ar-O}$~\cite{TYU13:134307,Mirahmadi2022}.

Since most of experimental data available are for M=Ar~\cite{LUT05:2764}, we worked on the reaction, O+O$_2$+Ar $\rightarrow$ O$_3$+Ar, to benchmark our theoretical approach. Note that in many laboratory experiments \cite{hippler1990temperature,rawlins1987dynamics,LUT05:2764,janssen2001kinetic}, it has been shown that the rate coefficient $k_3$ for the three-body recombination with M=N$_2$ have very similar behavior as those with M=Ar.

    Fig.~\ref{fig:k3T_QU} displays calculated thermally-averaged recombination rates of ozone molecules, scaled by the factor of 70~K$/T$, accounting for vibrational quenching of the highly-excited ozone molecules after three-body recombination. Furthermore, it is found that most O$_3$  molecules formed initially are highly exited; thus, taking into account the quenching is necessary. The scaling factor is a statistical factor that can be interpreted as the survival probability of highly excited ozone molecules (related to the ratio of stabilization events to the total number of events, i.e.,  stabilization and dissociation) formed through the direct three-body recombination. $70$~K comes from the order of the energy of Ar-O$_3$ interaction averaged over relevant vibrational states of O$_3$. It is important to note that the dependence $\propto 1/T$ is valid only for $T\le\Delta E$. For lower temperatures, the survival factor is $\approx1$. It is worth mentioning that Troe suggested the same dependence for the stabilization factor {\it et al.}~\cite{LUT05:2764,troe1977theory}.

   As it can be seen in Fig.~\ref{fig:k3T_QU}, the calculated values agree well with experimental data for temperatures between 100~K and 900~K. However, for the temperatures below $100$~K, the results do not match the experimental data. This can be viewed as a restriction on the applicability of a classical approach for these temperatures since, at the corresponding collision energies, details of the resonant structure of O$_3$ and, possibly, ArO$_2$ metastable rovibrational states could be substantial, which cannot be treated accurately using classical-trajectory methods.

     \begin{figure}
    	\centering
    	\includegraphics[scale=0.5]{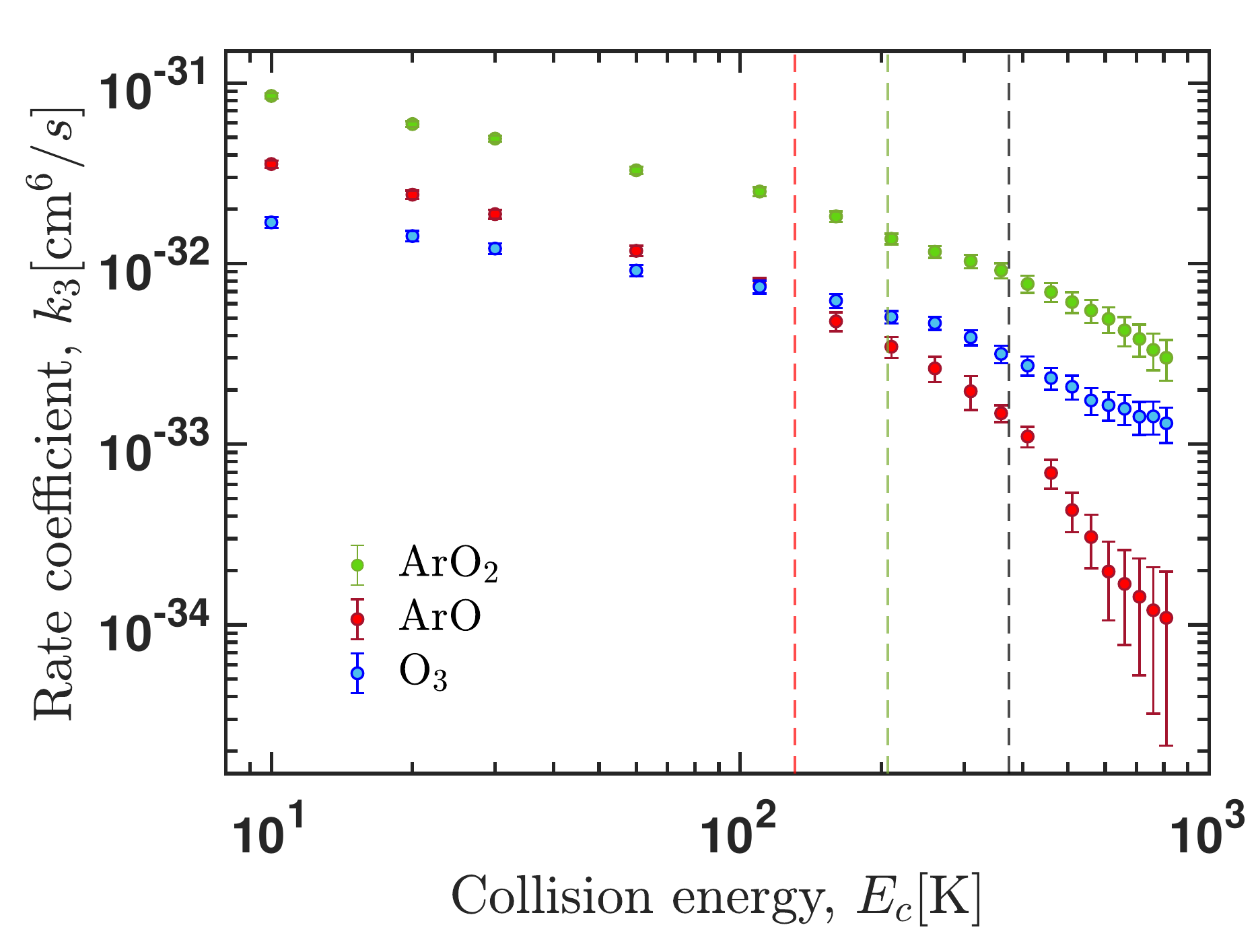}
    	\caption{Ternary recombination rate coefficient $k_3(E_c)$ for formation of the three possible products in the O+O$_2$+Ar collisions: O$_3$, ArO$_2$, and ArO. Red and green dashed lines indicate the dissociation energies of ArO and ArO$_2$, respectively. The black dashed line shows the energy related to shoulder structure of O-O$_2$. [Reproduced with permission from \cite{Mirahmadi2022}.] }
    	\label{fig:k3T_ALL}
    \end{figure}     
    
As exposed above, a direct approach of three-body recombination brings new insights into the relevance of energy-transfer and Chaperon mechanism for the ozone formation reaction. As shown in Fig.~\ref{fig:k3T_ALL}, it is possible to obtain all the reaction products after three-body recombination as a function of the collision energy. It is noticed that at low collision energies,$E_c\lesssim 200$~K,  ArO$_2$, and ArO prevail as the main reaction products, implying that the Chaperon mechanism is the proper one to explain ozone formation. On the contrary, at high collision energies, the O$_3$ shows a more significant rate than  ArO and a more gentle trend as a function of the collision energy than ArO$_2$ production. Hence, for $E_c\gtrsim 700$~K, the energy-transfer mechanism dominates the fate of the ozone formation reaction. However, for collision energies between 200~K and 700~K, ozone formation appears through a combination of both mechanisms, showing the new insights brought by a direct approach to three-body recombination reactions.

\subsubsection{Charged-neutral-neutral systems}\label{mionatomatom}

Molecular ion-atom-atom three-body recombination, or ion-molecule ternary association reactions, at cryogenic temperatures are highly relevant for rotational state action spectroscopy techniques for molecular ions. The reaction AB$^+$ + He + He$\rightarrow$ ABHe$^+$ + He is studied as a function of the internal state of the molecular ion driven by an external laser. Next, by looking into the state-selective reactivity, it is possible to find high-resolution spectra of the rotational structure of the molecule. Following this approach, the rotational structure of different molecular ions has been revealed. With it, the molecular ion-atom-atom three-body recombination is measured at cryogenic temperatures, as shown in Table \ref{tab1}. In the case of polyatomic ions, the internal structure of the molecular ion affects the three-body recombination rate, leading to rates up to 100 times larger than diatomic ions.

On the contrary, in the case of diatomic molecular ions, the three-body recombination rate remains unchanged primarily when different species are considered. In particular, the rate change is of the order of 10$\%$ or less. However, all the charged molecules studied show dipole moments, leading to a similar long-range attractive atom-ion interaction. Therefore, in virtue of capture models based on the long-range tail of the interaction, the discrepancy between polyatomic and diatomic ions should depend on the role of internal degrees of freedom.

	\begin{table}[t]
 \begin{center}
		\caption{\label{tab1} Molecule-He-He three-body recombination measured in ion traps in the presence of He buffer gas. Values taken from Ref.~\cite{BRUNKEN2017}. Here, $T$ represents the temperature associated to a two-body collision.  }
%		\begin{ruledtabular}
			\begin{tabular}{ccc}
				Molecule	& T(K) & $k_3(10^{-30} \text{cm}^6/\text{s})$  \\
				\hline
			CD$^+$ & 6 & 0.96 $\pm$ 0.11\\
			CO$^+$ & 5 & 0.9 $\pm$ 0.1\\
			CF$^+$ & 5 & 0.8 $\pm$ 0.1\\
			HCO$^+$ & 5 & 1.54 $\pm$ 0.09\\
			C$_3$H$^+$ & 5 & 0.5 $\pm$ 0.1\\
			NH$_3$D$^+$ & 6 & 1.5 $\pm$ 0.2\\
			CH$_2$D$^+$ & 6 & 4 $\pm$ 1\\
			CD$_2$H$^+$ & 6 & 2.8 $\pm$ 0.5\\
				\end{tabular}
%		\end{ruledtabular}
\end{center}
	\end{table}

\begin{table}[t]
\begin{center}
		\caption{\label{tab2} Molecule-atom-atom three-body recombination reactions observed in a temperature range between 82 and 287K, and characterized by $n$, following Eq.(\ref{eqk300n})  }
%		\begin{ruledtabular}
			\begin{tabular}{lcc}
				System	 & $n$  & Reference\\
				\hline
	C$^+$ + H$_2$ + He $\rightarrow$ CH$_2^+$ + He  & 1.3 &\cite{ADAMS1981}\\
	C$^+$ + D$_2$ + He $\rightarrow$ CD$_2^+$ + He  & 1.2 &\cite{ADAMS1981}\\
				\end{tabular}
%		\end{ruledtabular}
\end{center}
	\end{table}

On the other hand, theoretical investigations, based mainly on the Rice–Ramsperger–Kassel–Marcus (RRKM) theory~\cite{Troe1977,Bates1979,Bates1979bis,Bates1985,Bates1988,Herbst1979,Herbst1979bisbis,Herbst1980,Rainer1986,Viggiano1986,Troe2005}, show that the three-body recombination is given by

\begin{equation}
\label{eqk300n}
    k_3(T)=k_3(300\text{K})\left( \frac{T}{300\text{K}}\right)^{-n},
\end{equation}
where $k_3(300\text{K})$ represents the three-body recombination rate at 300~K. In the case of reactants with internal degrees of freedom, it has been proposed that $k_{3}(T)\propto T^{-l/2}$, where $l$ represents the number of rotational degrees of freedom of the reactants~\cite{Bates1979}. That is, $K_3(T)\propto T^{-1}$ for molecule-atom-atom collisions. However, this theoretical approach does not explain the experimental findings displayed in Table~\ref{tab2}, showing $k_{3}(T)\propto T^{-n}$, based on Eq.~(\ref{eqk300n}) with $n=1.2$ and $n=1.3$ for two different systems. Therefore, the role of internal degrees of freedom of the reactants is not fully captured by previous theoretical approaches based on the existence of an intermediate complex. In addition, it is worth noticing that in the case of Table~\ref{tab2}, one of the neutral species is the one with internal degrees of freedom, whereas in Table~\ref{tab1} is the charged species with the one showing internal degrees of freedom. In this arena, a direct approach based on classical trajectory calculations could bring a different perspective, as it has been shown in ozone formation~\cite{Mirahmadi2022}.

\subsection{Molecule-molecule-atom three-body recombination}\label{mma}

Molecule-molecule-atom three-body recombination reactions have been widely studied in the literature. However, in most scenarios, one of the reactants is charged. First, as displayed in Table~\ref{tab3}, we focus on the reported three-body recombination rates at a given temperature. This Table shows the rates for room temperature, except for H$^+$ + H$_2$  + H$_2$  $\rightarrow$ H$_3^+$ + H$_2$, displaying three-body recombination rates between $10^{-26}$ and $10^{-31}$~cm$^6$/s. Indeed, in some systems, the rate is as large such that the three-body recombination becomes more critical than two-body fragmentation channels~\cite{Smith1978}.

%However, some general conclusions about the recombination physics can be drawn from this data. 

\begin{table}[t!]
		\caption{\label{tab3} Molecule-molecule-atom three-body recombination measured at a given temperature. Reaction rates with $^*$ indicate a lower bound of the actual experimental rate.  }
%		\begin{ruledtabular}
			\begin{tabular}{lccc}
				System	& T(K) & $k_3(10^{-30} \text{cm}^6/\text{s})$ & Reference \\
				\hline
H$^+$ + H$_2$  + H$_2$  $\rightarrow$ H$_3^+$ + H$_2$ & 11 & 30$\pm10$ & \cite{zymak2011}\\
H$_3$O$^+$ + N$_2$ + He$\rightarrow$ H$_3$O$^+$N$_2$ + He  & 300 & 3.4 &\cite{Smith2021} \\
H$_3$O$^+$ + O$_2$ + He$\rightarrow$ H$_3$O$^+$O$_2$ + He  & 300 & 0.18 &\cite{Smith2021} \\
NO$^+$ + O$_2$ + He$\rightarrow$ NO$^+$O$_2$ + He  & 300 & 0.22 &\cite{Smith2021} \\
O$_2^+$ + O$_2$ + He$\rightarrow$ O$_4^+$ + He  & 300 & 0.12 &\cite{Smith2021} \\
H$_3$O$^+$ + CO$_2$ + He$\rightarrow$ H$_3$O$^+$CO$_2$ + He  & 300 & 60& \cite{Smith2021} \\
NO$^+$ + CO$_2$ + He$\rightarrow$ NO$^+$CO$_2$ + He  & 300 & 2.3 &\cite{Smith2021} \\
O$_2^+$ + CO$_2$ + He$\rightarrow$ O$_2^+$CO$_2$ + He  & 300 & 3.0 &\cite{Smith2021} \\
H$_3$O$^+$ + H$_2$O + He$\rightarrow$ H$_3$O$^+$H$_2$O + He  & 300 & 3300 &\cite{Smith2021} \\
NO$^+$ + H$_2$O + He$\rightarrow$ NO$^+$H$_2$O + He  & 300 & 220 &\cite{Smith2021} \\
O$_2^+$ + H$_2$O + He$\rightarrow$ O$_2^+$H$_2$O + He  & 300 & 450 &\cite{Smith2021} \\
CH$_3^+$ + H$_2$ + He$\rightarrow$ CH$_3^+$H$_2$ + He  & 300 & 130 &\cite{Smith1978} \\
CH$_3^+$ + N$_2$ + He$\rightarrow$ CH$_3^+$N$_2$ + He  & 300 & 53 &\cite{Smith1978} \\
CH$_3^+$ + O$_2$ + He$\rightarrow$ CH$_3^+$O$_2$ + He  & 300 & 10 &\cite{Smith1978} \\
CH$_3^+$ + CO + He$\rightarrow$ CH$_3^+$CO + He  & 300 & 2200 &\cite{Smith1978} \\
CH$_3^+$ + CO$_2$ + He$\rightarrow$ CH$_3^+$CO$_2$ + He  & 300 & 710 &\cite{Smith1978} \\
CH$_3^+$ + H$_2$O + He$\rightarrow$ CH$_3^+$H$_2$O + He  & 300 & 30000$^*$ &\cite{Smith1978} \\
CH$_3^+$ + NH$_3$ + He$\rightarrow$ CH$_3^+$NH$_3$ + He  & 300 & 70000$^*$ &\cite{Smith1978} \\
CH$_3^+$ + H$_2$CO + He$\rightarrow$ CH$_3^+$H$_2$CO + He  & 300 & 35000 &\cite{Smith1978} \\
CH$_3^+$ + CH$_3$OH + He$\rightarrow$ CH$_3^+$CH$_3$OH + He  & 300 &  40000$^*$&\cite{Smith1978} \\
CH$_3^+$ + CH$_3$NH$_2$ + He$\rightarrow$ CH$_3^+$CH$_3$NH$_2$ + He  & 300 & 3000$^*$ &\cite{Smith1978} \\
				\end{tabular}
%		\end{ruledtabular}
	\end{table}

The data shown in Table~\ref{tab3} helps draw some general behavior of the three-body recombination rate as a function of the colliding species. In the case of O$_2^+$ reactions, we notice that the reaction with H$_2$O shows the most significant rate, which can be rationalized by considering that H$_2$O has a permanent dipole moment. As a result, the dominant long-range interaction is charged-dipole, leading to a larger capture volume than charged-induced dipole interactions, characteristic of atom-ion systems. On the contrary, in the case of  O$_2^+$ + O$_2$ + He, and O$_2^+$ + CO$_2$ + He, using a capture model for charged-neutral interactions, one would expect a change of 20$\%$, at most, due the different polarizabilities of O$_2$ (1.56$\times 10^{-24}$~cm$^3$~\cite{CRC}) and CO$_2$ (2.91$\times 10^{-24}$~cm$^3$~\cite{CRC}). Instead, a change of one order of magnitude between the rates of these reactions is observed. Long-range interactions can not explain this change; thus, internal degrees of freedom must be relevant to the reaction dynamics. Similarly, by looking into O$_2^+$ + O$_2$ + He versus CH$_3^+$ + O$_2$ + He reactions, one notices that the second reaction shows a 100 times larger reaction rate than the first one, even though the neutrals are the same in both reactions. Hence, the long-range capture volume should be identical. Therefore, one can conclude that the internal degrees of freedom of the molecular ion play a significant role in the three-body recombination dynamics. 

However, to fully understand the role of internal degrees of freedom, it is better to look into the temperature dependence of the three-body recombination rate following Eq.~(\ref{eqk300n}) and displayed in Table~\ref{tab4}, showing a wide range of $n$ between 1 and 4. Some reactions like CH$_3^+$ + CO + He $\rightarrow$ CH$_3$CO$^+$ + He are well described within the RRKM theoretical framework. In contrast, others, like Cl$^-$ + H$_2$ + H$_2$ $\rightarrow$ ClH$_2^-$ +H$_2$, show a drastic deviation from the RRKM prediction. Hence, it is very convoluted to draw any general consideration about the RRKM theory approach and the role of internal degrees of freedom of the reactants. In this case, a direct approach to three-body recombination could enlighten the role of internal degrees of freedom of the colliding bodies.

Moreover, it is noticed that some reactions have been measured at different temperature ranges using various experimental methods, leading to different values for the temperature-dependent rate, as in the case of N$_2^+$ + N$_2$ + He $\rightarrow$ N$_4^+$ + He. Measurements based on drift tubes by B\"ohringer et al.~\cite{Bohringer1983}, and selected ion flow tube (SIFT) technique of Adams and Smith~\cite{ADAMS1976349} agree very well with each other, but they differ from the measurements performed in a drift source by van Koppen et al.~\cite{vanKoppen1984}. This disagreement was noticed in the work of van Koppen et al.~\cite{vanKoppen1984}, although it was not possible to explain. Thus, some of the reported temperature-dependent rates in the literature must be revisited. 

\begin{table}[t]
\begin{center}    
		\caption{\label{tab4}Molecule-molecule-atom three-body recombination reactions measured in a given range of temperatures, $T$ in K. $n$ denotes the power-dependence of the three-body recombination rate as a function of the temperature as given by Eq.~(\ref{eqk300n}).  }
%		\begin{ruledtabular}
			\begin{tabular}{lccc}
			System	& T (in K) & $n$  & Reference\\
				\hline
Cl$^-$ + H$_2$ + H$_2$ $\rightarrow$ ClH$_2^-$ +H$_2$ & 14-26 & 0.98$\pm 0.09$ & \cite{Wild2021} \\
%CH$_3^+$ + CO + He $\rightarrow$ CH$_3$CO$^+$ + He & 100-300 & 2.7 & \cite{Adams1979}\\
CH$_3^+$ + CO + He $\rightarrow$ CH$_3$CO$^+$ + He & 82-520 & 2.4 & \cite{ADAMS1981}\\		
%CH$_3^+$ + H$_2$ + He $\rightarrow$ CH$_3$H$_2^+$ + He & 100-300 & 4.4 & \cite{Adams1979}\\	 
CH$_3^+$ + H$_2$ + He $\rightarrow$ CH$_3$H$_2^+$ + He & 82-520 & 2.3 & \cite{ADAMS1981}\\		 
%CH$_3^+$ + N$_2$ + He $\rightarrow$ CH$_3$N$_2^+$ + He & 100-300 & 3.8 & \cite{Adams1979}\\
CH$_3^+$ + N$_2$ + He $\rightarrow$ CH$_3$N$_2^+$ + He & 82-520 & 2.7 & \cite{ADAMS1981}\\		 
%CH$_3^+$ + O$_2$ + He $\rightarrow$ CH$_3$O$_2^+$ + He & 100-300 & 4.0 & \cite{Adams1979}\\
CH$_3^+$ + O$_2$ + He $\rightarrow$ CH$_3$O$_2^+$ + He & 82-287 & 2.8 & \cite{ADAMS1981}\\		 
%CH$_3^+$ + CO$_2$ + He $\rightarrow$ CH$_3$CO$_2^+$ + He & 100-300 & 3.8 & \cite{Adams1979}\\
CH$_3^+$ + CO$_2$ + He $\rightarrow$ CH$_3$CO$_2^+$ + He & 200-520 & 3.4 & \cite{ADAMS1981}\\	CH$_3^+$ + H$_2$O + He $\rightarrow$ CH$_3$H$_2$O$^+$ + He & 367-800 & 3.3 & \cite{BARASSIN1980269}\\
CD$_3^+$ + D$_2$ + He $\rightarrow$ CD$_5^+$ + He & 82-520 & 2.5 & \cite{ADAMS1981}\\	
CD$_3^+$ + N$_2$ + He $\rightarrow$ CD$_3$N$_2^+$ + He & 82-520 & 2.7 & \cite{ADAMS1981}\\	
CD$_3^+$ + CO + He $\rightarrow$ CD$_3$CO$^+$ + He & 82-520 & 2.3 & \cite{ADAMS1981}\\	
CD$_3^+$ + O$_2$ + He $\rightarrow$ CD$_3$O$_2^+$ + He & 82-300 & 2.3 & \cite{ADAMS1981}\\		CD$_3^+$ + CO$_2$ + He $\rightarrow$ CD$_3$CO$_2^+$ + He & 200-520 & 3.3 & \cite{ADAMS1981}\\	
%N$_2^+$ + N$_2$ + He $\rightarrow$ N$_4^+$ + He & 82-520 & 2.4 & \cite{Adams1979bis}\\	
N$_2^+$ + N$_2$ + He $\rightarrow$ N$_4^+$ + He & 40-520 & 2.3$\pm 0.2$ & \cite{Bohringer1983} \\
N$_2^+$ + N$_2$ + He $\rightarrow$ N$_4^+$ + He & 80-520 & 2.4$\pm 0.2$ & \cite{ADAMS1981} \\
N$_2^+$ + N$_2$ + Ne $\rightarrow$ N$_4^+$ + Ne & 80-200 & 1.54$\pm 0.02$ & \cite{vanKoppen1984} \\
NO$_3^-$ + HCl + He $\rightarrow$ NO$_3$HCl$^-$ + He & 153-300 & 3.52 & \cite{Viggiano1984}\\
NO$_3^-$HNO$_3$ + HCl + He $\rightarrow$ NO$_3^-$HNO$_3$HCl + He & 153-300 & 3.68 & \cite{Viggiano1984}\\
N$^+$ + N$_2$ + N$_2$ $\rightarrow$ N$_3^+$ +N$_2$ & 120-440 & 2 & \cite{DHEANDHANOO1983} \\
N$^+$ + N$_2$ + N$_2$ $\rightarrow$ N$_3^+$ +N$_2$ & 220-500 & 2.1$\pm 0.2$ & \cite{Guthrie1991} \\	
CO$^+$ + CO + He $\rightarrow$ (CO)$_2^+$ + He & 80-530 & 1.22$\pm0.05$ & \cite{VANKOPPEN198641} \\
CO$^+$ + CO + He $\rightarrow$ (CO)$_2^+$ + He & 80-530 & 1.22$\pm0.05$ & \cite{VANKOPPEN198641} \\
CO$^+$ + CO + He $\rightarrow$ (CO)$_2^+$ + He & 80-700 & 1.5 & \cite{Meot1974} \\
CO$^+$ + CO + Ne $\rightarrow$ (CO)$_2^+$ + Ne & 80-530 & 1.20$\pm0.05$ & \cite{VANKOPPEN198641} \\
				\end{tabular}
\end{center}
\end{table}

\subsection{Molecule-molecule-molecule three-body recombination}\label{mmm}

\begin{table}[t]
 \begin{center}
		\caption{\label{tab5} Molecule-molecule-molecule three-body recombination reactions in a given range of temperatures, $T$ in K. $n$ denotes the power-dependence of the three-body recombination rate as a function of the temperature as given by Eq.~(\ref{eqk300n}). }
%		\begin{ruledtabular}
			\begin{tabular}{lccc}
			System	& T (in K) & $n$  & Reference\\
				\hline
NO$_3^-$ + HCl + H$_2$ $\rightarrow$ NO$_3^-$HCl + H$_2$ & 153-300 & 2.02 & \cite{Viggiano1984}\\
NO$_3^-$ + HCl + N$_2$ $\rightarrow$ NO$_3^-$HCl + N$_2$ & 153-300 & 2.62 & \cite{Viggiano1984}\\
NO$_3^-$HNO$_3$ + HCl + H$_2$ $\rightarrow$ NO$_3^-$HNO$_3$HCl + H$_2$ & 153-300 & 4.17 & \cite{Viggiano1984}\\
NO$_3^-$(HNO$_3$)$_2$ + HCl + H$_2$ $\rightarrow$ NO$_3^-$(HNO$_3$)$_2$HCl + H$_2$ & 153-300 & 5.59 & \cite{Viggiano1984}\\
O$_2^+$ + O$_2$ + O$_2$ $\rightarrow$ O$_4^+$ +O$_2$ & 51-340 & 2.93 & \cite{Bohringer1982}  \\
O$_2^+$ + O$_2$ + O$_2$ $\rightarrow$ O$_4^+$ +O$_2$ & 80-350 & 3.2 & \cite{Payzant1973}  \\
N$_2^+$ + N$_2$ + N$_2$ $\rightarrow$ N$_4^+$ + N$_2$ & 45-400 & 1.64 & \cite{Bohringer1982} \\
N$_2^+$ + N$_2$ + N$_2$ $\rightarrow$ N$_4^+$ + N$_2$ & 120-440 & 2.14 & \cite{DHEANDHANOO1983} \\
N$_2^+$ + N$_2$ + N$_2$ $\rightarrow$ N$_4^+$ + N$_2$ & 220-500 & 2.2 $\pm0.2$& \cite{Guthrie1991} \\
N$_2^+$ + N$_2$ + N$_2$ $\rightarrow$ N$_4^+$ + N$_2$ & 250-700 & 1.7 & \cite{Meot1974} \\
N$_2^+$ + N$_2$ + N$_2$ $\rightarrow$ N$_4^+$ + N$_2$ & 80-455 & 1.67$\pm 0.07$ & \cite{vanKoppen1984}\\
N$_2^+$ + N$_2$ + N$_2$ $\rightarrow$ N$_4^+$ + N$_2$ & 20-160 & 1.85$\pm 0.07$ & \cite{Rowe1984} \\
N$_2^+$ + N$_2$ + N$_2$ $\rightarrow$ N$_4^+$ + N$_2$ & 5-15 & 2.05$\pm 0.05$ & \cite{Randeniya1989} \\
O$_2^+$ + N$_2$ + N$_2$ $\rightarrow$ O$_2$N$_2^+$ +N$_2$ & 100-180 & 3.2 & \cite{DHEANDHANOO1983} \\
NO$^+$ + N$_2$ + N$_2$ $\rightarrow$ NO$_2$N$_2^+$ +N$_2$ & 100-180 & 4.3 & \cite{DHEANDHANOO1983} \\
NO$^+$ + NO + NO $\rightarrow$ NO$_2^+$ +NO & 200-430 & 2.4$\pm0.1$ & \cite{VANKOPPEN198641} \\
NO$^+$ + CO$_2$ + N$_2$ $\rightarrow$ NO$^+$CO$_2$ +N$_2$ & 100-180 & 4.0 & \cite{DHEANDHANOO1983} \\
NO$^+$ + NO + NO $\rightarrow$ (NO)$_2^+$ +NO$_2$ & 200-430 & 1.6$\pm0.1$ & \cite{VANKOPPEN198641} \\      
CO$^+$ + CO + CO $\rightarrow$ (CO)$_2^+$ +CO$_2$ & 120-650 & 1.5 & \cite{Meot1974} \\
CO$^+$ + CO + CO $\rightarrow$ (CO)$_2^+$ +CO$_2$ & 350-550 & 1.6$\pm0.1$ & \cite{VANKOPPEN198641} \\
CO$^+$ + CO + CO $\rightarrow$ (CO)$_2^+$ +CO$_2$ & 220-500 & 2.4$\pm0.1$ & \cite{Guthrie1991}\\
CO$^+$ + CO + CO $\rightarrow$ (CO)$_2^+$ +CO$_2$ & 65-300 & 1.6 & \cite{Schlemmer2002}\\
(CO)$_2^+$ + CO + CO $\rightarrow$ (CO)$_3^+$ +CO$_2$ & 80-220 & 3.2 & \cite{Schlemmer2002}
    \end{tabular}
%		\end{ruledtabular}
\end{center}
\end{table}

The ultimate three-body recombination process is the molecule-molecule-molecule three-body recombination, in which all the reactants have internal degrees of freedom. Several of these reactions have been reported in the literature. A summary of those is displayed in Table~\ref{tab5}, showing the explored temperature range and the value of $n$ by fitting the experimental data to Eq.~(\ref{eqk300n}). First, as in the previous section, one observes a broad range of $n$ values ranging from 1.5 to 5.59. However, in this case, most of the reported values correspond to the same reaction studied at different temperature ranges and distinct experimental techniques, particularly, N$_2^+$ + N$_2$ + N$_2$ $\rightarrow$ N$_4^+$ + N$_2$ and CO$^+$ + CO + CO $\rightarrow$ (CO)$_2^+$ +CO$_2$ reactions. Different groups have extensively studied these reactions using other experimental techniques at different temperatures.

The N$_2^+$ + N$_2$ + N$_2$ $\rightarrow$ N$_4^+$ + N$_2$ shows a value of $n$ between 1.64 and 2.2. On the other hand, another study has studied this reaction for a fixed temperature but for a wide range of pressures~\cite{B105886J}. Thus, this reaction has enough experimental information to be theoretically modeled. In that endeavor, the work of Troe using a capture model with a falloff curve correction leads to the best result~\cite{Troe2005}. However, the results depend on the pressure since the proposed reaction mechanism relies on an indirect approach.

The CO$^+$ + CO + CO $\rightarrow$ (CO)$_2^+$ +CO$_2$ reaction has been studied at different temperatures leading to a wide range of $n$-values between 1.5 and 2.4. However, this reaction has been studied with modern trapping and imaging techniques. The ions are trapped in a Paul trap and then monitored as a buffer gas is added into the system, which leads to $n=1.6$. Surprisingly enough, three of the reported measurements agree on $n=1.6$. In contrast, a fourth one gives a much larger $ n$ value. Thus, that measurement should be revisited. On the theory side, only the RRKM theory has been developed for these reactions, which leads to a value of $n=2$, somehow in agreement with the reported coefficients. 

However, at this point, there is no general theory about three-body recombination, including reactants with internal degrees of freedom. Moreover, the available model relies on the existence of intermediate complex and statistical properties of this rather than in a first principles explanation due to the complexity of the problem. However, a direct approach to these scenarios will bring a new perspective and hints about the main reaction mechanisms behind molecule-molecule-molecule three-body recombination.

\section{Outlook and future challenges}\label{outlook}

Three-body recombination processes, or ternary association reactions, are termolecular reactions relevant to many systems in physics and chemistry in a wide range of density and temperature: from ultracold atoms and cold chemistry to nuclear fusion reactors. In physical chemistry, three-body recombination was systematically studied in the '70s, including molecule-molecule-atom and molecule-molecule-molecule processes. At the same time, theoretical models based on the Lindemann–Hinshelwood mechanism, or indirect approach, were derived and applied. Since then, three-body recombination reactions have been studied from an indirect approach, although only finding an accurate description of the reaction dynamics in particular cases. However, it is possible to describe three-body recombination reactions from a direct approach, i.e., without invoking the existence of an intermediate complex, thus, incorporating all possible reaction mechanisms associated with a given reaction. 

Here, we review a theoretical approach for direct three-body recombination, without invoking the existence of an intermediate complex, based on classical trajectory calculations in hyperspherical coordinates. This approach has proven successful in ion-atom-atom three-body recombination in the cold regime, predicting that molecular ions are the main reaction products. Accordingly, the three-body recombination rate depends on the temperature as $T^{-3/4}$. Both of these predictions have been experimentally corroborated in hybrid ion-atom trap experiments. In cryogenic environments, the same method predicts that any van der Waals molecule appears at almost the same rate in a buffer gas cell. Thus, nearly any X-He molecule, where X is any atom, could be formed in a buffer gas cell. In more complex three-body reactions, like ozone formation reaction, a direct approach has been proven to lead to good results compared to experimental data. In addition, we have been able to find the range of validity of the energy-transfer and Chaperon reaction mechanism from first principles. Moreover, we have developed a capture model for three-body collisions leading to threshold laws that have been experimentally proved in the case of ion-atom-atom three-body recombination. The same approach has been applied to N-body collisions, finding that the hyperradial distribution follows a universal trend. In other words, the shape of the hyperradial distribution function is independent of the details of the potential, collision energy, and mass of the colliding particles. 

Most of the experimental data available refer to molecule-atom-atom, molecule-molecule-atom, and molecule-molecule-molecule three-body recombination, in which a direct approach still needs to be developed. In addition, a theoretical framework based on the indirect approach fails to properly consider the internal degrees of freedom of the interacting bodies. Hence, further development on direct three-body recombination methods, including internal degrees of freedom, could elucidate the role of the internal degrees of freedom in three-body recombination, a long-standing problem in physical chemistry.

Despite their interest and relevance, three-body recombination reactions have received less attention than they deserve. However, with the development of cryogenic ion traps for spectroscopy purposes, it is possible to study, more and better than ever, three-body recombination reactions more accurately. Moreover, it could be possible to study the role of the initial quantum state of the reactants and analyze the product states. Therefore, a general three-body approach, including degrees of freedom, could be timely, bringing a new era for theoretically and experimentally three-body processes.

\section*{Acknowledgements}

M. Mirahmadi acknowledges the support of the Deutsche Forschungsgemeinschaft (DFG – German Research Foundation) under the grant number PE 3477/2 - 493725479. J. P.-R. acknowledges the support of the Simons Foundation.

%I would like to thank to Prof. G. Meijer for his support, Prof. C. H. Greene for being my mentor, Prof. S. Willitsch for this opportunity and to my collaborators that have made this possible. In addition, I would like to thank M. T. Cretu and H. Hirzler for carefully reading the manuscript and for useful suggestions.  

\section*{Disclosure statement}
No potential conflict was reported by the auhtors.

\bibliographystyle{tfo}
\bibliography{3BR}

\end{document}